\documentclass[acmsmall,authorversion,nonacm]{acmart}
\usepackage{microtype}[draft]
\usepackage{xcolor}
\usepackage{xspace}
\usepackage{tikz}
\usepackage{listings}
\usepackage{cprotect}
\usepackage{enumitem}
\setlist[itemize]{noitemsep, topsep=0pt, leftmargin=*}

\usetikzlibrary{automata,positioning}

\title{Hinted Dictionaries: Efficient Functional Ordered Sets and Maps}

\author{Amir Shaikhha}
\affiliation{
  \institution{University of Edinburgh}            
  \country{United Kingdom}                    
}

\author{Mahdi Ghorbani}
\affiliation{
  \institution{University of Edinburgh}            
  \country{United Kingdom}                    
}

\author{Hesam Shahrokhi}
\affiliation{
  \institution{University of Edinburgh}            
  \country{United Kingdom}                    
}

\ccsdesc[500]{Software and its engineering~Functional languages}
\ccsdesc[500]{Theory of computation~Data structures design and analysis}

\keywords{Functional Collections, Ordered Dictionaries, Sparse Linear Algebra.}

\begin{document}

\newcommand{\codespace}{\vspace{0.2cm}}
\colorlet{revisioncolor}{black}
\colorlet{revisionbgcolor}{lightgray!30!white}
\newcommand{\revision}[1]{#1}

\newcommand{\straightquote}[0]{\textquotesingle\!\!\textquotesingle}
\newcommand{\codeColor}[0]{teal!90!black} 

\newcommand{\smartpara}[1]{\vspace{.3em}\noindent \textbf{#1.}}

\lstdefinelanguage{Scala}%
{morekeywords={abstract,%
  case,catch,char,class,%
  def,else,extends,final,finally,for,%
  if,import,implicit,%
  match,module,%
  new,null,%
  object,override,%
  package,private,protected,public,%
  for,public,return,super,%
  this,throw,trait,try,type,%
  val,var,%
  with,while,do,%
  yield, sealed,
  },%
  sensitive,%
  morecomment=[l]//,%
  morecomment=[s]{/*}{*/},%
  morestring=[b]",%
  morestring=[b]',%
  showstringspaces=false,%
  moredelim=**[is][\$\color{black}]{~}{~},%
  moredelim=**[is][\color{black}]{^}{^},
  moredelim=**[is][\color{\codeColor}]{⌜}{⌝},
}[keywords,comments,strings]%

\lstset{language=Scala,%
  backgroundcolor = \color{lightgray!30!white},
  escapeinside={(*@}{@*)},%
  breaklines=true,%
  mathescape=true,%
showspaces=false,
showtabs=false,
showstringspaces=false,
breakatwhitespace=true,
  aboveskip=1pt,
  belowskip=1pt,
  lineskip=-0.2pt,
  frame=single,
   numbersep=5pt,
   numberstyle=\small\ttfamily,
   basicstyle=\small\ttfamily,
  keywordstyle=\small\ttfamily\bfseries,%
columns=fullflexible,
  commentstyle=\color{gray},
  emph={%
     foreach
    },emphstyle={\color{blue}\bfseries},
  escapeinside={(*@}{@*)},
  literate=%
  {⟨}{{\color{black}}$\langle${\color{\codeColor}}}1
  {⟩}{{\color{black}}$\rangle${\color{black}}}1
  {‹}{{\color{black}}\color{\codeColor}\textbf{code}\!\straightquote}5
  {›}{{\color{\codeColor}\straightquote}\color{black}}1
  {⌈}{{\color{black}\{}}1
  {⌉}{\}{\color{\codeColor}}}1
  {€}{\$}1 
}
\newcommand{\code}[1]{\lstinline[
  language=Scala,
  breaklines=true,
  showspaces=false,
  showtabs=false,
  showstringspaces=false,
  breakatwhitespace=true,
  numbers=none,
  numberstyle=\ttfamily,
  basicstyle=\ttfamily,
  keywordstyle=\ttfamily\bfseries,%
  columns=fullflexible,
  escapeinside={(*@}{@*)}    
]|#1|}

\lstset{language=Scala} 

\newcommand{\todo}[1]{\textcolor{red}{\tiny{\bf TODO:} \footnotesize#1}}

\newcommand{\amir}[1]{\textcolor{red}{\tiny{\bf Amir:} \footnotesize#1}}

\newcommand{\LPchange}[2]{#2}

\newcommand{\cutForSpace}[1]{}
\newcommand{\cutForLaterMaybe}[1]{}

\newcommand{\paper}{article\xspace}


\begin{abstract}
This \paper introduces hinted dictionaries for expressing efficient ordered sets and maps functionally. As opposed to the traditional ordered dictionaries with logarithmic operations, hinted dictionaries can achieve better performance by using cursor-like objects referred to as hints. Hinted dictionaries unify the interfaces of imperative ordered dictionaries (e.g., C++ maps) and functional ones (e.g., Adams' sets). We show that such dictionaries can use sorted arrays, unbalanced trees, and balanced trees as their underlying representations. Throughout the \paper, we use Scala to present the different components of hinted dictionaries. We also provide a C++ implementation to evaluate the effectiveness of hinted dictionaries. Hinted dictionaries provide superior performance for set-set operations in comparison with the standard library of C++. Also, they show a competitive performance in comparison with the SciPy library for sparse vector operations.
\end{abstract}

\maketitle

\section{Introduction}
Sets and maps are two essential collection types for programming used widely in data analytics~\cite{sdql}.
The underlying implementation for both \revision{are normally based on 1)} hash tables or \revision{2)} ordered data structures. \revision{The former provides (average-case) constant-time lookup, insertion, and deletion operations, while the latter performs these operations in a logarithmic time.}
The trade-off between these two approaches has been heavily investigated in systems communities~\cite{kim2009sort}.

An important class of operations are those dealing with two collection types, such as set-set-union or the merge of two maps. 
One of the main advantages of hash-based implementations is a straightforward implementation for such operations with a linear computational complexity. 
\revision{However, na\"ively using ordered dictionaries results in an implementation with a computational complexity of $O(n \log(n))$.}

\revision{\smartpara{Motivating Example}}
\revision{The} following C++ code computes the intersection of two sets, implemented by \code{std::unordered_set}, a hash-table-based set:

\begin{lstlisting}[language=C++]
std::unordered_set<K> result;
for(auto& e : set1) {
  if(set2.count(e))
  	result.emplace(e);
}
\end{lstlisting}

However, the same fact is not true for ordered data structures; changing the dictionary type to \code{std::set}, an ordered implementation, results in a program with \revision{$O(n \log(n))$} computational complexity. This is because both the \code{count} (lookup) and \code{emplace} (insertion) methods have logarithmic computational complexity.

As a partial remedy, the standard library of C++ provides an alternative insertion method that can take linear time, if used appropriately. 
The \code{emplace\_hint} method takes a hint for the position that the element will be inserted. 
If the hint correctly specifies the insertion point, the computational complexity will be amortized to constant time.\footnote{https://www.cplusplus.com/reference/set/set/emplace\_hint/}

\begin{lstlisting}[language=C++]
std::set<K> result;
auto hint = result.begin();
for(auto& e : set1) {
  if(set2.count(e))
  	hint = result.emplace_hint(hint, e);
}
\end{lstlisting}

However, the above implementation still suffers from an \revision{$O(n \log(n))$} computational 
complexity\revision{, due to the logarithmic computational complexity of the lookup operation (\code{count}) of the second set.
Thanks to the orderedness of the second set, one can observe that once an element is looked up, 
there is no longer any need to search its preceding elements at the next iterations.
By leveraging this feature, we can provide a \textit{hinted} lookup method with an amortized constant run-time.

\smartpara{Hinted Data Structures} The following code, shows an alternative implementation for set intersection that uses such hinted lookup operations:}

\begin{lstlisting}[language=C++,backgroundcolor=\color{revisionbgcolor}]
hinted_set<K> result;
hinted_set<K>::hint_t hint = result.begin();
for(auto& e : set1) {
  hinted_set<K>::hint_t hint2 = set2.seek(e);
  if(hint2.found)
    hint = result.insert_hint(hint, e);
  set2.after(hint2);
}
\end{lstlisting}

\revision{
The above \textit{hinted set} data-structure enables faster insertion and lookup by providing a cursor through a \textit{hint object} (of type \code{hint\_t}).
The \code{seek} method returns the hint object \code{hint2} pointing to element \code{e}. 
Thanks to the invocation of \code{set2.after(hint2)}, the irrelevant elements of \code{set2} (which are smaller than \code{e}) are no longer considered in the next iterations. 
The expression \code{hint2.found} specifies if the element exists in \code{set2} or not.
Finally, if an element exists in the second set (specified by \code{hint2.found}), it is inserted into its correct position using \code{insert\_hint}.
}

\revision{This \paper introduces \textit{hinted dictionaries, a class of functional ordered data structures}.
The essential building block of hinted dictionaries are \textit{hint objects}, that enable faster operations (than the traditional $O(\log n)$ complexity) by maintaining a pointer into the data structure.

\smartpara{Related Work}
The existing work on efficient ordered dictionaries can be divided into two categories. First, in the imperative world, there are C++ ordered dictionaries (e.g., \code{std::map}) with \textit{limited} hinting capabilities only for insertion through \code{emplace\_hint}, but not for deletion and lookup, as observed previously. 
}

\revision{Second, from the functional world, Adams' sets~\cite{adams1993functional} provide efficient implementations for set-set operators.}
Functional languages such as Haskell have implemented ordered sets and maps based on them for more than twenty years~\cite{haskellmap}. 
Furthermore, it has been shown~\cite{blelloch2016just} that Adams' \revision{maps} can be used to provide a parallel implementation for balanced trees such as AVL~\cite{avltree}, Red-Black~\cite{rbtree}, Weight-Balanced~\cite{wbtree}, and Treaps~\cite{treaptree}.
\revision{However, Adams' maps do not expose any hint-based operations to the programmer. 
At first glance, these two approaches seem completely irrelevant to each other. 
}

\smartpara{Contributions}
\textit{The key contribution of this \paper is hinted dictionaries, an ordered data structure that unifies the techniques from both imperative and functional worlds.}
The underlying representation for hinted dictionaries can be sorted arrays, unbalanced trees, and balanced trees by sharing the same interface.
In our running example, alternative data-structure implementations can be easily provided by simply changing the type signature of the hinted set from \code{hinted\_set} to another implementation, without modifying anything else. 

This \paper is organized as follows:

\begin{itemize}
\item We present monoid dictionaries, the most general form of dictionaries without any orderedness constraint on keys (Section~\ref{sec:monoiddict}). Such dictionaries subsume sets and maps and provide a restricted form of iterations in the form of map-reduce for computing associative and commutative aggregations over them (Section~\ref{sec:dictiter}). 
\item Afterwards, we show ordered dictionaries, a more restricted class of dictionaries where the keys need to be ordered (Section~\ref{sec:ordereddict}). The iterations over these dictionaries are more general than monoid dictionaries, by relaxing the commutative requirement and providing associative aggregations. We show two particular interfaces for implementing associative aggregations in Section~\ref{sec:ordereddictiter}.
\item We introduce hinted dictionaries, an implementation technique for ordered dictionaries (Section~\ref{sec:hinteddict}). 
A key ingredient of hinted dictionaries are the hint objects. In Section~\ref{sec:hintop} we show operations over hint objects.
\item Hinted dictionaries provide both sequential and parallel implementations for associative aggregations (Section~\ref{sec:hinteddictiter}). 
\item In order to support binary operations over dictionaries, hinted dictionaries provide a bulk operation interface (Section~\ref{sec:bulk}).
The design choice for these bulk operations results in a completely different instantiation of hinted dictionaries. 
We present two implementations in this \paper: insert-based (Section~\ref{sec:hinteddictinsert}) and join-based (Section~\ref{sec:hinteddictjoin}) hinted dictionaries.
\item We present the implementation for hint objects in Section~\ref{sec:hintimpl}. More specifically, we present \textit{focus-based hints}, hint objects focusing on a particular position in the dictionary (Section~\ref{sec:focushint}), and the corresponding focus-based hinted dictionaries (Section~\ref{sec:focushinteddict}).
\item We present the concrete implementations for hinted dictionaries in Section~\ref{sec:concrete}. More specifically, we show the implementation of ordered dictionaries using sorted arrays (Section~\ref{sec:sortedarray}), unbalanced trees (Section~\ref{sec:unbalancedtree}), and balanced trees (Section~\ref{sec:balancedtree}). 
\revision{Then, we connect all the components together in Section~\ref{sec:putit}.}
\item Finally, we provide a C++ prototype for hinted dictionaries and discuss the challenges for tuning its performance in Section~\ref{sec:tuned}.
We compare its performance with the standard library of C++ for set-set and sparse vector operations and show its asymptotic improvements in Section~\ref{sec:exp_res}. For the latter workload we show its competitive performance with SciPy.
\end{itemize}

Throughout the \paper, we use Scala \revision{to present the different components.}
\revision{The main ideas behind hinted dictionaries, however, can be implemented in other object-oriented and functional languages with support for generic types and} lambda expressions (e.g., Haskell, OCaml, Java, Julia, Rust, and C++), as demonstrated in Section~\ref{sec:tuned}.
\revision{
The former is required for implementing the structural interfaces (e.g., \code{Monoid[T]}) and generic dictionaries (e.g., \code{Dict[K, V]}), while the latter is essential for implementing higher-order functions (e.g., \code{mapReduce}) in hinted dictionaries.
}
\section{Monoid Dictionary}
\label{sec:monoiddict}

In this section, we present the most general form of  dictionaries that we support; the ones where the values form a monoid structure, referred to as monoid dictionaries. 
We start by defining monoid and equalable values. 
Afterwards, we introduce the interface for monoid dictionaries.
Finally, we show the class of iterations that can be expressed over them.

\subsection{Monoid}
\label{sec:monoid}
A monoid is defined as a set of values $V$, with a binary operator \code{op} and a \code{zero} element, such that the following properties hold:

\begin{itemize}
\item \textbf{Associativity:} For all elements $a$, $b$, $c$ in $V$: 
\code{op}(\code{op}($a$, $b$), $c$) = \code{op}($a$, \code{op}($b$, $c$))  
\item \textbf{Identity Element:} For all elements $a$ in $V$: 
\code{op}($a$, \code{zero}) = \code{op}(\code{zero}, $a$) = $a$
\end{itemize}

An important class of monoids, supports the following additional axiom:

\begin{itemize}
\item \textbf{Commutativity:} For all elements $a$, $b$ in $V$: 
\code{op}($a$, $b$) = \code{op}($b$, $a$)
\end{itemize}

\noindent Such monoids are referred to as \textit{commutative monoids}. Important examples of commutative monoids are boolean values under conjunction and disjunction, integer numbers under multiplication and addition. Matrices of real numbers are commutative monoids under addition, but are non-commutative monoids under multiplication.

The Scala interface for monoid structures is shown in Figure~\ref{fig:monoid}. This interface can be thought of as a type class, where providing concrete implementations for this interface results in type class instances.

\subsection{Equalable}
\label{sec:equalable}
In order to perform lookups over dictionaries, one requires to check for the equality of keys. 
This is achieved by the \code{Equalable} type class (Figure~\ref{fig:monoid}). 
Each type class instance provides the implementation strategy for checking the equivalence between two keys by overriding the \code{equiv} method.

\begin{figure}[t]
\begin{lstlisting}
trait Monoid[V] {
  def op(e1: V, e2: V): V
  def zero: V
}

trait Equalable[K] {
  def equiv(e1: K, e2: K): Boolean
}
\end{lstlisting}
\caption{The interface for monoid and equalable type classes.}
\label{fig:monoid}
\end{figure}

\subsection{Dictionary Interface}
\label{sec:dictinterface}
Given the key type \code{K} with an \code{Equalable[K]} constraint, and the value type \code{V} with a \code{Monoid[V]} constraint, one can define the interface \code{Dict[K, V, D]} for a dictionary type \code{D} (Figure~\ref{fig:dict}).

\begin{figure}[t]
\begin{lstlisting}[backgroundcolor=\color{revisionbgcolor}]
trait Dict[K, V, D] {
  implicit val equ: Equalable[K]
  implicit val mon: Monoid[V]
  def find(dict: D, key: K): V
  def insert(dict: D, key: K, value: V): D
  def delete(dict: D, key: K): D
  def size(dict: D): Int
  def count(dict: D): Int
  def empty(): D
  def isEmpty(dict: D): Boolean
  final def single(key: K, value: V): D = insert(empty(), key, value)
}
\end{lstlisting}
\caption{The interface for (monoid) dictionaries.}
\label{fig:dict}
\end{figure}

The specification of the methods of monoid dictionaries is as follows:

\begin{itemize}
\item \code{find}: performs a look up for the associated value with \code{key} in the dictionary \code{dict}. If the key does not exist in the dictionary, the identity element of the monoid structure over \code{V} is returned (\code{mon.zero}).
\item \code{insert}: first performs a look up for the associated value with \code{key}. If the key does not exist, the pair of \code{key} and \code{value} is inserted in the dictionary. If the key already exists, the associated value, \code{old_value}, is updated by applying the  binary operator of monoid to \code{old_value} and \code{value} (\code{mon.op(old_value, value)}). As the result, the updated dictionary is returned.
\item \code{delete}: returns a new dictionary where \code{key} and its associated value is removed.
\item \code{size}: returns the number of key-value pairs in the dictionary.
\revision{\item \code{count}: returns the number of key-value pairs with non-zero values in the dictionary.}
\item \code{empty}: returns an empty dictionary of type \code{D} with keys and values of type \code{K} and \code{V}.
\item \code{isEmpty}: check if the given dictionary is empty or not.
\item \code{single}: returns a singleton dictionary containing the pair of \code{key} and \code{value}. This can be implemented by inserting into an empty dictionary.
\end{itemize}

By providing appropriate monoid structures for values, one can instantiate different collections from monoid dictionaries. 
As an example, using boolean values results in sets, using natural numbers results in bags, and using \code{Option} types results in maps.\footnote{Using \code{Option} types incurs boxing and unboxing costs that is avoided by libraries such as \texttt{scala-unboxed-option}~\cite{unboxedoptions} for Scala and unpacked sums in GHC for Haskell~\cite{unpackedsums} }

\subsection{Iterations over Dictionaries}
\label{sec:dictiter}
Next, we introduce the constructs required for performing iterations over dictionaries.
As monoid dictionaries do not enforce any order over the keys, the iterative computation over them must be order-agnostic.
Otherwise, the iterative computation results in different outcomes depending on the underlying organization of dictionary keys.

We provide the \code{mapReduce} method for expressing sound iterations over monoid dictionaries.
This method performs map-reduce operations by starting from the initial element \code{z}, computing a transformation between key-value pairs to element of result type by \code{map}, and reducing the result elements by \code{red}. 
To ensure the soundness of aggregate computations, the \code{red} binary operator must be both commutative and associative. 

As an alternative interface, we provide \code{aggregate} with a monoid \revision{constraint} over the result type. 
This method can be implemented in terms of \code{mapReduce} (cf. Figure~\ref{fig:dictiter}).
\revision{To do so, we need to use the zero element and the binary operator of an instance of the type class \code{Monoid[R]}. 
The Scala type system can provide an instance for type class \code{T} by using \code{implicitly[T]}~\cite{oliveira2010type}.
In this case, \code{implicitly[Monoid[R]]} returns an instance of type \code{Monoid[R]}, where its zero} element \revision{(\code{monR.zero})} and binary operator \revision{(\code{monR.op})} are \revision{passed} as the initial value and reduction functions of \code{mapReduce}\revision{, respectively}.
Note that for sound aggregations, the result type should form a commutative monoid.

One can easily provide an implementation for \code{size} using the \code{aggregate} method. As we are only interested in counting the number of key-value pairs, it is sufficient to transform them to \code{1}. \revision{Note that \code{size} returns the number of all pairs in the dictionary, while \code{count} only returns the number of pairs containing non-zero values.}

\revision{For a cleaner presentation, we use the \code{DictIteration} interface to define the iteration-based methods (cf. Figure~\ref{fig:dictiter}). 
This way, we avoid a large interface for \code{Dict}.
To do so, we need to make sure that all classes and interfaces that extend \code{DictIteration}, also extend the \code{Dict} interface. 
This is achieved by ascribing the type of \code{this} object of the \code{DictIteration} interfaces with \code{Dict[K, V, D]}.
Such dependency injection technique is known as \textit{cake pattern} in the Scala programming language.
It is important to note that using this technique is not essential;
we can implement this code in a language without this feature by removing the interface for \code{DictIteration} altogether. Instead, all the method definitions of \code{DictIteration} are added to \code{Dict}.}

\begin{figure}[t]
\begin{lstlisting}[backgroundcolor=\color{revisionbgcolor}]
trait DictIteration[K, V, D] { this: Dict[K, V, D] =>
  // precond: `red` must be commutative and associative
  def mapReduce[R](dict: D, z: R, map: (K, V) => R, red: (R, R) => R): R
  // precond: `R` must form a commutative monoid
  def aggregate[R: Monoid](dict: D, map: (K, V) => R): R = {
    val monR = implicitly[Monoid[R]]
    mapReduce[R](dict, monR.zero, map, monR.op)
  }
  def size(dict: D): Int = 
    aggregate[Int](dict, (k, v) => 1)
  def count(dict: D): Int = 
    aggregate[Int](dict, (k, v) => if(v == mon.zero) 0 else 1)
}
\end{lstlisting}
\caption{The interface for iterations on dictionaries.}
\label{fig:dictiter}
\end{figure}

\section{Ordered Dictionary}
\label{sec:ordereddict}
In this section we present ordered dictionaries, the keys of which should follow a total order. 
First, we define the required interface for orderable keys in Section~\ref{sec:orderable}. 
Then, we introduce the interface for ordered dictionaries including bulk operations of them in Section~\ref{sec:ordereddictinterface}.
Finally, similar to monoid dictionaries, we show the class of iterations expressible over ordered dictionaries in Section~\ref{sec:ordereddictiter}.

\subsection{Orderable}
\label{sec:orderable}

In this section, we introduce the interface required for the keys of ordered dictionaries (cf. Figure~\ref{fig:orderable}). 
In ordered dictionaries, apart from the need to check for the equality of two keys (using \code{equiv} derived from \code{Equalable}), a total order \revision{must also} be provided. 

The \code{compare} method is sufficient to provide the total order information.
If its return value is a positive number, the first element is greater than the second value, and for a negative number, vice versa.
Otherwise, if the return value is zero, this means that both elements are equal.
All comparison operators can be implemented using the \code{compare} method, as can be seen in Figure~\ref{fig:orderable}.

As we are only interested in finite dictionaries, one can provide an upper bound and lower bound for keys.
These values are specified using \code{max} and \code{min} in the \code{Orderable} interface.
We will see in Section~\ref{sec:focushint} how upper bounds can be used for implementing hint objects.

\begin{figure}[t]
\begin{lstlisting}
trait Orderable[K] extends Equalable[K] {
  def compare(e1: K, e2: K): Int
  def max: K
  def min: K
  final def lt(e1: K, e2: K): Boolean = compare(e1, e2) < 0
  final def gt(e1: K, e2: K): Boolean = compare(e1, e2) > 0
  final def lteq(e1: K, e2: K): Boolean = compare(e1, e2) <= 0
  final def gteq(e1: K, e2: K): Boolean = compare(e1, e2) >= 0
  final def equiv(e1: K, e2: K): Boolean = compare(e1, e2) == 0
}
\end{lstlisting}
\caption{The interface for orderable.}
\label{fig:orderable}
\end{figure}

\subsection{Ordered Dictionary Interface}
\label{sec:ordereddictinterface}
The interface of ordered dictionaries is very similar to monoid dictionaries.
The \code{Equalable} type class instance is the same as the one used for \code{Orderable}. 
This is because the \code{Orderable} interface subsumes the interface of \code{Equalable} by using inheritance.

The additional methods provided for ordered dictionaries are as follows:

\begin{itemize} 
\item \code{toList}: this method converts ordered dictionaries into a list of type \code{List[(K, V)]}.\footnote{Unordered dictionaries cannot implement \code{toList} as there is no deterministic order for the key-value pairs.}
\item \code{append}: for two ordered dictionaries \code{left} and \code{right}, where all the keys of \code{left} are less than the keys of \code{right}, this method returns an ordered dictionary containing the elements of both dictionaries.
\item \code{join}: given two ordered dictionaries \code{left} and \code{right} and a key-value pair \code{key} and \code{value}, this method creates an ordered dictionary containing the elements of \code{left} and \code{right} as well as the pair of \code{key} and \code{value}. \revision{In this method, all the keys of \code{left} must be less than \code{key}, and all the keys of \code{right} must be more than \code{key}.}
\end{itemize}

\revision{Note that the mentioned preconditions for the last two methods are necessary to preserve the dictionary's order, and violating any of them makes hinted dictionaries not work. Furthermore, these two} methods are bulk operations and thus are defined in a separate \code{OrderedDictBulkOps} interface (Figure~\ref{fig:ordereddict}) following the cake pattern. These methods are critical for providing different concrete ordered dictionary implementations, as will be observed in Section~\ref{sec:bulk}.

\begin{figure}[t]
\begin{lstlisting}
trait OrderedDict[K, V, D] extends Dict[K, V, D] {
  implicit val ord: Orderable[K]
  implicit val equ: Equalable[K] = ord
  def toList(dict: D): List[(K, V)]
}

trait OrderedDictBulkOps[K, V, D] { this: OrderedDict[K, V, D] =>
  // precond: keys(left) < keys(right)
  def append(left: D, right: D): D
  // precond: keys(left) < k < keys(right)
  def join(left: D, key: K, value: V, right: D): D
}
\end{lstlisting}
\caption{The interface for ordered dictionaries and bulk operations over them.}
\label{fig:ordereddict}
\end{figure}

\subsection{Iterations over Ordered Dictionaries}
\label{sec:ordereddictiter}
As opposed to monoid dictionaries, ordered dictionaries do not need the reduction function to be commutative. 
This is because even with non-commutative reductions, ordered dictionaries will result in a deterministic order for key-value pairs.

\begin{figure}[t]
\begin{lstlisting}
trait OrderedDictIteration[K, V, D] extends DictIteration[K, V, D] { 
  this: OrderedDict[K, V, D] =>
  // precond: `red` should only be associative 
  def mapReduce[R](dict: D, z: R, map: (K, V) => R, red: (R, R) => R): R
  def toList(dict: D): List[(K, V)] = 
    aggregate[List[(K, V)]](dict, (k, v) => List((k, v)))
}

trait OrderedDictFoldLeft[K, V, D] extends OrderedDictIteration[K, V, D] { 
  this: OrderedDict[K, V, D] =>
  def foldLeft[R](dict: D, z: R, op: (R, K, V) => R): R
  override def mapReduce[R](dict: D, z: R, map: (K, V) => R, red: (R, R) => R): R = 
    foldLeft[R](dict, z, (s, k, v) => red(s, map(k, v)))
}

trait OrderedDictFoldTree[K, V, D] extends OrderedDictIteration[K, V, D] { 
  this: OrderedDict[K, V, D] =>
  def foldTree[R, M](dict: D, z: R, op: (K, V, R) => (R, R, M), 
                        comb: (K, V, M, R, R) => R): R
  override def mapReduce[R](dict: D, z: R, map: (K, V) => R, red: (R, R) => R): R = 
    foldTree[R, Unit](dict, z, (k, v, s) => (s, s, ()), 
                        (k, v, _, s1, s2) => red(red(s1, map(k, v)), s2))
}
\end{lstlisting}
\caption{The interface for iterations over ordered dictionaries. The \code{foldLeft} method corresponds to computing aggregations sequentially and \code{foldTree} method is a parallel-friendly aggregate computation strategy.}
\label{fig:ordereddict:iter}
\end{figure}

The \code{toList} method can be implemented by \code{mapReduce} and \code{aggregate} methods. 
Figure~\ref{fig:ordereddict:iter} shows its implementation using \code{aggregate}; it is sufficient to map each of the key-value pairs into a singleton list. This requires the following instance of \code{Monoid[List[T]]}:

\begin{lstlisting}
  implicit def MonoidList[T] = new Monoid[List[T]] {
    def op(e1: List[T], e2: List[T]): List[T] = e1 ++ e2
    def zero: List[T] = Nil
  }
\end{lstlisting}

\noindent Here, the binary operator is list append (\code{++}) and the zero element is the empty list (\code{Nil}).
The \code{toList} method returns the result of appending all these singleton lists. 

\subsubsection{Sequential Iterations}
Similar to functional lists in functional languages, ordered dictionaries also provide a \code{foldLeft} method for performing accumulating computations over their elements \revision{from left to right}.
This method is provided in the \code{OrderedDictFoldLeft} interface (cf. Figure~\ref{fig:ordereddict:iter}).

The \code{mapReduce} method can be implemented using \code{foldLeft} by passing the initial value and defining the accumulating function as applying the \code{red} function to the previous state \code{s} and the result of \code{map(k, v)}.

\subsubsection{Parallel Iterations}
Thanks to the associative nature of reduction functions, there is no need to perform aggregations only sequentially from left to right. 
Instead, one can perform aggregations in a tree-structured manner, which is more parallel-friendly. 

The \code{foldTree} method provided in the \code{OrderedDictFoldTree} interface (cf. Figure~\ref{fig:ordereddict:iter}) is responsible for computing parallel iterations. 
This is provided by performing a top down traversal over the logical tree representation.
Similar to \code{foldLeft}, this method accepts an initial state (\code{z}). 
At each stage, it computes the stage to be passed to each of the subtrees. The \code{op} method produces a triple of elements when applied to the current key-value pair and the previous state.
The first two elements of this triple are the states to be passed to each of subtrees.
The last element corresponds to a hidden state of type \code{M}.
This hidden state is used after the aggregation for subtrees are computed.
The \code{comb} method applies this hidden state alongside with the current key-value and the states return by the subtrees and computes the next state. 

The \code{mapReduce} method can be implemented using \code{foldTree} as well. As the \code{op} method, we return the previous state \code{s} to both subtrees. 
As the \code{comb} method, we apply the reduction method twice. The first application involves the return state of left subtree (\code{s1}) and the mapping of key-value pair \code{map(k, v)}. The second application is over the result of the previous reduction and the state of the right subtree (\code{s2}).

Note that for implementing \code{mapReduce} there was no hidden state required, and thus the unit value \code{()} with unit type \code{Unit} was provided. 
We will see cases where this hidden state will be required in Section~\ref{sec:setset}.

By carefully keeping the value of aggregation in the ordered dictionary, one can provide a more efficient implementation for \code{mapReduce}.
This is similar to the idea of augmented trees, and has already been investigated in Augmented Maps~\cite{sun2018pam}.

\section{Hinted Dictionary}
\label{sec:hinteddict}
In this section, we introduce hinted dictionaries, an implementation strategy for ordered dictionaries. 
First, we present the interface of hinted dictionaries in Section~\ref{sec:hinteddictinterface}.
Then, we show the interface for hint objects in Section~\ref{sec:hintop}.

\begin{figure}[t]
\begin{lstlisting}
trait HintedDict[K, V, D, H] extends OrderedDict[K, V, D] {
  def begin(dict: D): H
  def middle(dict: D): H
  def end(dict: D): H
  def isEnd(dict: D, hint: H): Boolean
  def next(dict: D, hint: H): H
  def seek(dict: D, key: K): H
  // precond: current(hint)._1 == key
  def findHint(dict: D, hint: H, key: K): V
  // precond: current(hint)._1 == key
  def insertHint(dict: D, hint: H, key: K, value: V): D
  // precond: current(hint)._1 == key
  def deleteHint(dict: D, hint: H, key: K): D
  def insert(dict: D, key: K, value: V): D = 
    insertHint(dict, seek(dict, key), key, value)
  def delete(dict: D, key: K): D = 
    deleteHint(dict, seek(dict, key), key)
  def find(dict: D, key: K): V = 
    findHint(dict, seek(dict, key), key)
}
\end{lstlisting}
\caption{The interface for hinted dictionaries.}
\label{fig:hintdict}
\end{figure}

\subsection{Hinted Dictionary Interface}
\label{sec:hinteddictinterface}
Hinted dictionaries \revision{inherit} all the methods of \revision{both monoid dictionaries and ordered dictionaries}. 
Additionally, they provide the following methods:

\begin{itemize}
\item \code{begin}: returns the hint object corresponding to the beginning of dictionary. This method is useful for accessing the head of an ordered dictionary.
\item \code{middle}: returns the hint object of the middle of the dictionary. This method is useful for cases that require viewing the dictionary as a tree. For example, it can be used for a binary search where one needs to access the middle of a collection.
\item \code{end}: returns the hint object specifying the end of dictionaries.
\item \code{isEnd}: checks whether the given hint object corresponds to the end of the dictionary. This is useful for terminating an iteration over the hinted dictionary.
\item \code{next}: returns the hint object succeeding the given hint object over the input dictionary.
\item \code{seek}: returns the hint object for the position in the array where \code{key} would be placed. This means that the preceding elements have smaller keys and succeeding elements have larger keys.
In the case that the dictionary contains the \code{key}, the corresponding hint object is returned.
\item \code{findHint}: returns the associated value with the given key using the provided hint object. As the precondition, the hint should point to the correct position. 
\item \code{insertHint}: inserts the given key-value pair to the position provided by the hint object. Similar to the previous method, the hint object assumed to point to the correct position.
\item \code{deleteHint}: deletes the key-value pair corresponding to the input key using the provided hint object. A similar precondition to the previous two methods hold. 
\end{itemize}

By using the \code{seek} method to compute the correct hint object, one can have an implementation for \code{find}, \code{insert}, and \code{delete} using the corresponding hinted methods.
Once hinted dictionaries are supplied with (amortized) constant-time operations for these hinted methods, one can better benefit from their efficiency.

\begin{figure}[t!]
\centering
\includegraphics[width=\columnwidth]{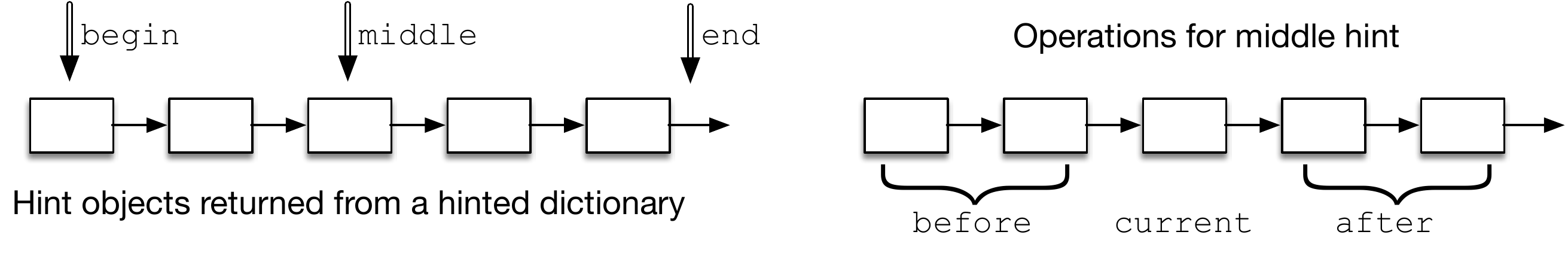}
\caption{An example hinted dictionary representation showing the hint objects returned by \code{begin}, \code{middle}, and \code{end} methods and the outcome of \code{before}, \code{current}, and \code{after} over the hint object returned by \code{middle}.}
\label{fig:hintobjfig}
\end{figure}
\begin{figure}[t]
\begin{lstlisting}[backgroundcolor=\color{revisionbgcolor}]
trait HintOps[K, V, D, H] { this: HintedDict[K, V, D, H] =>
  def before(hint: H): D
  def after(hint: H): D
  def current(hint: H): (K, V)
}
\end{lstlisting}
\caption{The interface for operations on hint objects.}
\label{fig:hintops}
\end{figure}

\subsection{Hint Operations}
\label{sec:hintop}
The hint objects can be thought as pointers to different locations of an ordered dictionary (cf. Figure~\ref{fig:hintobjfig}).
Rather than providing a concrete implementation for hint object, we provide an interface for them in Figure~\ref{fig:hintops}. We leave the actual implementation for these operators to Section~\ref{sec:hintimpl}.

The methods for hint objects are as follows:

\begin{itemize}
\item \code{before}: returns the dictionary of elements located before the hint object.
\item \code{after}: returns the dictionary of elements that are after the hint object.
\item \code{current}: returns the key-value pair of the entry of the dictionary that hint object is pointing into. If for a given key, there is no associated value, the identity element of monoid is returned, similar to what was observed for \code{find}.
\end{itemize}

As it was shown in Figure~\ref{fig:hintdict} we can use these methods to provide preconditions for the hinted dictionary methods. In addition, all the implementations of the \code{seek} method need to enforce the following post-condition. Assuming the result hint object is \code{res}, the key of this element should be the same as the input key (\code{current(res)._1 == key}).\revision{\footnote{\revision{Note that tuple indexing in Scala starts from 1, and \code{tup._i} where \code{i} is an integer shows the \code{i}th element of the tuple.}}} Furthermore, the keys of the dictionary before the result hint object (\code{before(res)}) should be less than \code{key}. Similarly, the keys of the dictionary after the result hint object (\code{after(res)}) should be greater than \code{key}.

\section{Iterations}
\label{sec:hinteddictiter}

In this section, we use the methods provided by hinted dictionaries to implement sequential and parallel iterations over them.

\begin{figure}[t]
\begin{lstlisting}[backgroundcolor=\color{revisionbgcolor}]
trait HintedDictFoldLeft[K, V, D, H] extends OrderedDictFoldLeft[K, V, D] 
  with HintedDict[K, V, D, H] with HintOps[K, V, D, H] {
  def foldLeft[R](dict: D, z: R, op: (R, K, V) => R): R = {
    @tailrec def foldLeftTR(hint: H, res: R): R =
      if(!isEnd(dict, hint)){
        val (key, value) = current(hint)
        val next_res = op(res, key, value)
        val next_hint = next(dict, hint)
        foldLeftTR(next_hint, next_res)
      }
      else
        res
    foldLeftTR(begin(dict), z)
  }
}
\end{lstlisting}
\caption{The implementation of sequential (fold-left-based) iterations over hinted dictionaries.}
\label{fig:hdictfoldl}
\end{figure}

\subsection{Sequential Implementation}
\revision{The interface of \code{HintedDictFoldLeft} containing the implementation for \code{foldLeft} is presented in Figure~\ref{fig:hdictfoldl}. 
Recall that this method is useful for stateful iterations over dictionaries. 
The initial state is specified by \code{z}, and the state is updated by applying the function \code{op} to the current state and key-value pairs.

In order to implement \code{foldLeft}, we define a recursive function \code{foldLeftTR} inside it.
This function has two input parameters \code{hint} and \code{res}, which specify the current hint object
and the computed state by iterating up to that hint object.}
We start by \revision{passing} the initial state \code{z} \revision{as} the value of \code{res}, and the beginning of the dictionary \revision{as} the value of \code{hint}.
\revision{The function \code{foldLeftTR} is recursively called until}
the hint object does not point to the end of dictionary (\code{!isEnd(dict, hint)})\revision{.
At each recursive call, 
we compute the next state value (\code{next_res})
}
by applying \code{op} to the current \revision{state (\code{res})} and key-value pair
\revision{and the next hint object (\code{next_hint})} by \code{next(dict, hint)}.

\revision{Note that the definition of \code{foldLeftTR} is annotated with \code{@tailrec}.
This means that this function is tail recursive -- all recursive calls are appearing as the last statement.
This annotation ensures that the Scala compiler can turn this method into imperative \code{while} loops, which results in better performance by removing the need to increase the stack frame size. 
}

\subsubsection{Example: Sparse Vector Inner Product}
Figure~\ref{fig:sparsevec} shows the interface for sparse vectors.
A sparse vector \code{Vec} can be represented as a dictionary from indices to a scalar value \code{Sca}.
In order to support operation such as inner product that involve multiplication over scalar values, we need to
define a type class instance for monoid over scalar values under multiplication (specified by \code{prod}).

An efficient sequential implementation of the inner product is provided in Figure~\ref{fig:sparsevec}.
The \code{foldLeft} method accepts a pair of states containing the result of inner product as well as the rest of the second vector to process. 
The state is initially set to the monoid identity element (\code{dict.mon.zero}) and the second vector (\code{v2}).
At each iteration, the \code{seek} method for the given key is invoked over the remaining part of the second vector.
Then, the result of inner product is updated by adding the previous result (\code{res}) to the outcome of multiplying (\code{prod.op}) the current value (\code{v}) with the value specified by the hint object (\code{current(hint)._2}). 
Note that in the case that the second vector does not have any elements at index \code{k}, the specified value by its hint object would be zero (\code{dict.mon.zero}).
Finally, the rest of the second vector is computed by taking only the elements of dictionary happening after the hint object (\code{after(hint)}).

\begin{figure}[t]
\begin{lstlisting}
trait SparseVectorOps[Sca, Vec] {
  implicit val prod: Monoid[Sca]
  def inner(v1: Vec, v2: Vec): Sca
}

trait SparseVectorFoldLeftOps[Sca, Vec, H] extends SparseVectorOps[Sca, Vec] {
  val dict: HintedDict[Int, Sca, Vec, H]
  val dictFolding: OrderedDictFoldLeft[Int, Sca, Vec]
  val hintOps: HintOps[Int, Sca, Vec, H]
  import dict._
  import dictFolding._
  import hintOps._
  def inner(v1: Vec, v2: Vec): Sca =
    foldLeft[(Sca, Vec)](v1, (dict.mon.zero, v2), (s, k, v) => {
      val (res, v2p) = s
      val hint = seek(v2p, k)
      dict.mon.op(res, prod.op(v, current(hint)._2)) -> after(hint)
    })._1
}
\end{lstlisting}
\caption{The implementation of the inner product of two sparse vectors using sequential iteration.}
\label{fig:sparsevec}
\end{figure}

\subsection{Parallel Implementation}
\begin{figure}[t!]
\centering
\includegraphics[width=0.7\columnwidth]{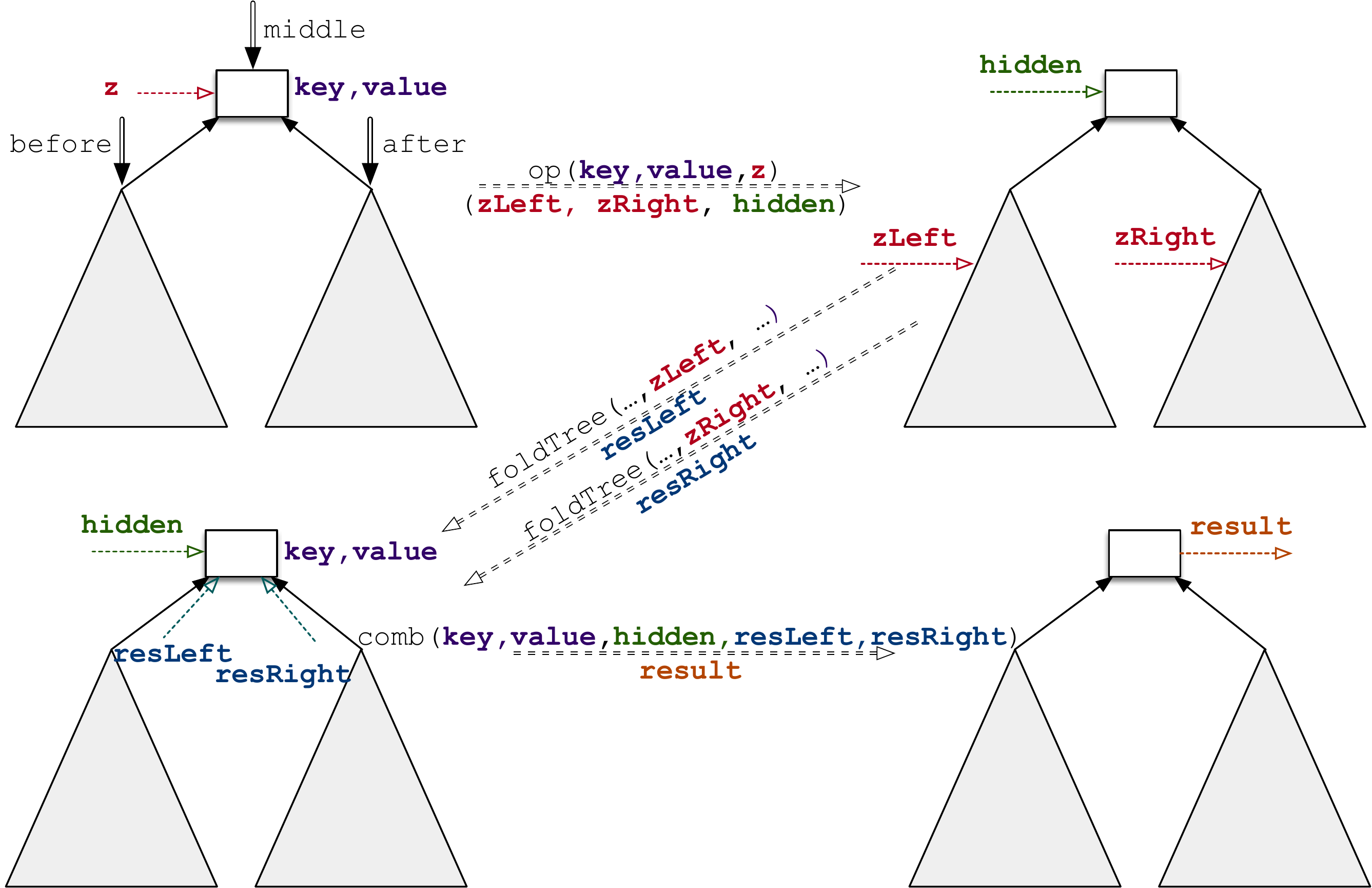}
\caption{\revision{The process of executing \code{foldTree} on a logical tree view of an ordered dictionary. The recursive invocations of \code{foldTree} on the two sub-trees can be evaluated in parallel.}}
\label{fig:hdictfoldtree}
\end{figure}

\revision{
The \code{foldTree} method can also be used for performing stateful iterations over hinted dictionaries. 
As opposed to \code{foldLeft}, this method can perform the computation in a divide-and-conquer manner.
Figure~\ref{fig:hdictfoldtree} demonstrates the process of applying \code{foldTree} over a hinted dictionary viewed as a tree.
The divide phase involves recursively applying \code{foldTree} to the left and right subtrees, where the initial state for each recursive call is computed using the function \code{op}.
The conquer phase uses the function \code{comb} to combine the results of recursive calls to compute the output state of the entire tree.
}

\revision{The interface of \code{HintedDictFoldTree} provides the implementation for \code{foldTree} using hinted dictionaries (cf. Figure~\ref{fig:hdictfoldt}).}
If the input dictionary is empty the initial value is returned. 
Otherwise, the \revision{following steps are performed}. First, the hint object returned by \code{middle} is used to retrieve the key-value pairs in the middle of the dictionary. 
The \code{op} function is applied to this key-value pair and the 
previous state \code{z}.
This results in the initial state of the left and right sub-trees (\code{zLeft} and \code{zRight}) as well as the hidden state (\code{hidden}).
The \code{foldTree} method is recursively invoked for both left and right sub-trees (\code{before(hint)} and \code{after(hint)}) using their corresponding initial states. These two invocations are independent of each other and can be run in parallel.
Finally, the key-value pair, the hidden state, and the results of recursive calls (\code{resLeft} and \code{resRight}) are combined by applying the \code{comb} function.

\begin{figure}[t]
\begin{lstlisting}
trait HintedDictFoldTree[K, V, D, H] extends OrderedDictFoldTree[K, V, D] 
  with HintedDict[K, V, D, H] with HintOps[K, V, D, H] {
  def foldTree[R, M](dict: D, z: R, op: (K, V, R) => (R, R, M), 
                      comb: (K, V, M, R, R) => R): R =
    if(isEmpty(dict)) z
    else {
      val hint = middle(dict)
      val (key, value) = current(hint)
      val (zLeft, zRight, hidden) = op(key, value, z)
      val resLeft = foldTree(before(hint), zLeft, op, comb)
      val resRight = foldTree(after(hint), zRight, op, comb)
      comb(key, value, hidden, resLeft, resRight)
    }
}
\end{lstlisting}
\caption{The implementation of parallel (fold-tree-based) iterations over hinted dictionaries.}
\label{fig:hdictfoldt}
\end{figure}

Next, we show example usages of these iteration constructs.

\subsubsection{Example: Set-Set Union}
\label{sec:setset}
A set of type \code{S}, consisting elements of type \code{K} can be expressed \revision{as} a dictionary with keys of type \code{K} and values of type \code{Boolean}.
Hence, they can also be expressed using hinted dictionaries.

Figure~\ref{fig:setset} provides the interface for set-set operations such as \code{union}, \code{intersect}, and \code{difference}.
For the sake of brevity here we only show the implementation for \code{union} using \code{foldTree}.
The \code{foldLeft}-based implementation and the one for \code{intersect} and \code{difference} can be similarly obtained. 

The result type of \code{foldTree} \revision{(cf. Figure~\ref{fig:setset})} is the set type \code{S}, and the type of its hidden state is the hint type \code{H}.
The iteration is over \code{set1} with the initial state of \code{set2}. 
At each stage, the \code{seek} method looks for the key \code{k} in the set specified by its current state \code{s}.
As the state for the left and right sub-trees, the dictionaries with preceding and succeeding elements (\code{before(hint)} and \code{after(hint)}) are provided, and as the hidden state the hint object \code{hint} is returned.
For the combine operation, the state returned by the left sub-tree (\code{s1}), the current key-value pair with update value (\code{k} and \code{mon.op(v, current(hidden)._2)}), and the state returned by the right sub-tree (\code{s2}) are joined.


\begin{figure}[t]
\begin{lstlisting}
trait SetSetOps[S] {
  def union(set1: S, set2: S): S
  def intersect(set1: S, set2: S): S
  def difference(set1: S, set2: S): S
}

trait SetSetFoldTreeOps[K, S, H] extends SetSetOps[S] {
  val dict: HintedDict[K, Boolean, S, H]
  val dictFolding: OrderedDictFoldTree[K, Boolean, S]
  val hintOps: HintOps[K, Boolean, S, H]
  val dictDictOps: OrderedDictBulkOps[K, Boolean, S]
  import dict._
  import dictFolding._
  import hintOps._
  import dictDictOps._

  def union(set1: S, set2: S): S = 
    foldTree[S, H](set1, set2, (k, v, s) => {
      val hint = seek(s, k)
      (before(hint), after(hint), hint)
    }, (k, v, hidden, s1, s2) => join(s1, k, mon.op(v, current(hidden)._2), s2))
  // ...
}
\end{lstlisting}
\caption{The implementation of set-set union using fold-tree-based iteration.}
\label{fig:setset}
\end{figure}

\section{Bulk Operations}
\label{sec:bulk}
In this section, we provide two design decisions for implementing bulk operations.
The first design is insert-based hinted dictionaries, where the bulk operations are derived from the implementation for the hinted insertion method (Section~\ref{sec:hinteddictinsert}).
The second design revolves around using the join method as the central operator (Section~\ref{sec:hinteddictjoin}).

\subsection{Insert-Based Hinted Dictionaries}
\label{sec:hinteddictinsert}
The hinted insertion method (\code{insertHint}) is an expressive operation. 
By providing a concrete implementation for this method, one can guide how an ordered dictionary can be inductively constructed starting from an empty dictionary.

Figure~\ref{fig:insertdict} shows the implementation of insert-based hinted dictionaries. 
The \code{insertHint} method is left as abstract. 
Providing different implementations result in a completely different strategy for maintaining ordered dictionaries.

The \code{append} method is implemented by iterating over the second dictionary and adding its elements one-by-one to the first dictionary. 
This is achieved by \code{foldLeft} over \code{right}, with an initial state of \code{left}, and adding the key-value pairs of \code{right} to the end of the previously computed result.
This implementation is correct because we know that the keys appearing in the second dictionary are greater than the keys of the first dictionary.
By assuming that we provide an implementation for \code{insertHint} with an amortized constant-time complexity,
the \code{append} operation will have an amortized linear run-time complexity.

The \code{join} method can be implemented in terms of \code{append} and \code{insertHint}. First, we need to insert the given key-value pair to the end of the first dictionary. Afterwards, this intermediate dictionary is appended by the second dictionary.
By making the same assumptions as \code{append}, this operation has also an amortized linear run-time complexity.

\begin{figure}[t]
\begin{lstlisting}
trait InsertBasedDict[K, V, D, H] extends HintedDict[K, V, D, H] 
  with HintOps[K, V, D, H] with OrderedDictBulkOps[K, V, D] 
  with HintedDictFoldLeft[K, V, D, H] {
  // precond: keys(left) < keys(right)
  def append(left: D, right: D): D = 
    foldLeft[D](right, left, (s, k, v) =>
      insertHint(s, end(s), k, v)
    )
  // precond: keys(left) < key < keys(right)
  def join(left: D, key: K, value: V, right: D): D = 
    append(insertHint(left, end(left), key, value), right)
  // precond: current(hint)._1 == key
  def insertHint(dict: D, hint: H, key: K, value: V): D
}
\end{lstlisting}
\caption{The implementation of insert-based hinted dictionaries.}
\label{fig:insertdict}
\end{figure}

\subsection{Join-Based Hinted Dictionaries}
\label{sec:hinteddictjoin}
An alternative way for defining hinted dictionaries is based on the join operator.
This is inspired by \revision{Adams'} sets~\cite{adams1993functional} and the follow up parallel implementations~\cite{blelloch2016just}.

The implementation of join-based hinted dictionaries is shown in Figure~\ref{fig:joindict}.
The \code{join} method does not have a concrete implementation.
It has been shown~\cite{blelloch2016just} that different balanced tree representations such as AVL~\cite{avltree}, Red-Black~\cite{rbtree}, Weight-Balanced~\cite{wbtree}, and Treaps~\cite{treaptree} 
can be expressed by providing an appropriate implementation for the \code{join} method.

The \code{append} method is expressed as follows. 
If the second dictionary is empty, the first dictionary is returned.
Otherwise, the hint object for the beginning of the second dictionary is used to retrieve its first key-value pair.
Then, the \code{join} method is applied to the first dictionary, the first key-value pair, and the rest of the second dictionary.
If the \code{join} method is an amortized linear operation, the \code{append} method also follows the same run-time complexity.

The \code{insertHint} method is expressed by joining the dictionary preceding the hint object (\code{before(hint)}), the key-value pair with updated value (\code{mon.op(current(hint)._2, value)}), and the dictionary succeeding the hint object (\code{after(hint)}).
Note that this way of implementing \code{insertHint} is suboptimal given that the \code{join} is a linear time operator. 
Thus, one has to try avoid using \code{insertHint} for join-based hinted dictionaries for performance reasons.

\begin{figure}[t]
\begin{lstlisting}
trait JoinBasedDict[K, V, D, H] extends HintedDict[K, V, D, H] 
  with HintOps[K, V, D, H] with OrderedDictBulkOps[K, V, D] {
  // precond: keys(left) < keys(right)
  def append(left: D, right: D): D = {
    if (isEmpty(right))
      left
    else {
      val hint = begin(right)
      val (key, value) = current(hint)
      val rightNew = after(hint)
      join(left, key, value, rightNew)
    }
  }
  // precond: keys(left) < key < keys(right)
  def join(left: D, key: K, value: V, right: D): D
  // precond: current(hint)._1 == key
  def insertHint(dict: D, hint: H, key: K, value: V): D = 
    join(before(hint), key, mon.op(current(hint)._2, value), after(hint))
}
\end{lstlisting}
\caption{The implementation of join-based hinted dictionaries.}
\label{fig:joindict}
\end{figure}

The efficiency of hinted dictionaries is not solely dependent on efficient \code{join} and \code{insertHint} operations.
Having an efficient hint object implementation is also essential, which will be presented next.

\section{Hint Implementation}
\label{sec:hintimpl}
This section starts with a concrete representation for hint objects. Using this representation we provide the implementation for hint operations in Section~\ref{sec:focushint}.
Afterwards, we provide the implementation of the rest of the methods of hinted dictionaries in Section~\ref{sec:focushinteddict}.

\subsection{Focus-Based Hints}
\label{sec:focushint}
As it was stated in Section~\ref{sec:hintop}, hint objects can be viewed as pointers to different locations of an ordered dictionary.
In this section, we consider them as objects focusing on a particular position in the dictionary.
The key-value pair that the hint object focuses on, specifies the \code{key} and \code{value} fields of the \code{FocusHint} class.
The lack of a key-value pair is specified by putting the identity element of the underlying monoid for type \code{V}.
The sub-dictionary containing the elements preceding/succeeding the focused key-value pair are stored in \code{left}/\code{right}.

The implementation of the hint operations using focus-based hints is very natural: \code{before} and \code{after} return \code{left} and \code{right} fields of the \code{FocusHint} object, and \code{current} returns the pair \code{(hint.key, hint.value)}.

\begin{figure}[t]
\begin{lstlisting}
case class FocusHint[K, V, D](left: D, key: K, value: V, right: D)

trait FocusHintOps[K, V, D] extends HintOps[K, V, D, FocusHint[K, V, D]] {
  this: HintedDict[K, V, D, FocusHint[K, V, D]] =>
  type H = FocusHint[K, V, D]
  def before(hint: H): D = hint.left
  def after(hint: H): D = hint.right
  def current(hint: H): (K, V) = (hint.key, hint.value)
}
\end{lstlisting}
\caption{The implementation for focus hint and its operations.}
\label{fig:focushint}
\end{figure}

\subsection{Focus-Based Hinted Dictionaries}
\label{sec:focushinteddict}
Figure~\ref{fig:focusdict} shows the implementation of focus-based hinted dictionaries, where the hint objects are \code{FocusHint}s.
The following methods are left as abstract: \code{empty}, \code{isEmpty}, \code{begin}, and \code{middle}. 
Depending on the underlying data structure, the implementation for these methods can be different.

The methods implemented by focus-based dictionaries are as follows:

\begin{itemize}
	\item \code{seek}: If the given dictionary is empty, an empty \code{FocusHint} object is returned the key of which is the same as the input key. Otherwise, it performs a binary search to return an appropriate hint object. 
	For binary search, the key of the middle of the dictionary is compared with the input key.
	If the middle key is the same as the input key, the middle hint object is returned. 
	If the input key is less than the middle key, the process is recursively invoked for the preceding dictionary (\code{seek(l, key)}), and the result hint object is computed by substituting its right dictionary by joining it with the rest of the input dictionary. 
	A similar process is performed when the input key is greater than the middle key.
	\item \code{end}: Returns an empty \code{FocusHint} object with the key set to the upper bound of keys (\code{ord.max}).
	\item \code{isEnd}: Check if the given hint is the same as the end hint object. 
	\item \code{next}: The hint object for the succeeding dictionary is constructed. If this hint object corresponds to the end of that dictionary, then the end of the input dictionary is returned. Otherwise, the return hint object is constructed by using this hint object and the preceding dictionary of the input dictionary.
	\item \code{deleteHint}: It is computed by appending the preceding and succeeding dictionaries together.
	\item \code{findHint}: The associated value with the hint object is returned. In the case where the key does not exist, the hint object stores the identity of the underlying monoid.
\end{itemize}

\begin{figure}[t!]
\begin{lstlisting}
trait FocusHintedDict[K, V, D] extends HintedDict[K, V, D, FocusHint[K, V, D]] 
  with FocusHintOps[K, V, D] with OrderedDictBulkOps[K, V, D] {
  // postcond: res.key == key
  def seek(dict: D, key: K): H = {
    if(isEmpty(dict))
      FocusHint(empty(), key, mon.zero, empty())
    else {
      val hint@FocusHint(l, m, v, r) = middle(dict)
      if(ord.equiv(key, m)) {
        hint
      } else if(ord.lt(key, m)) {
        val hint2 = seek(l, key)
        hint2.copy(right = join(hint2.right, m, v, r))
      } else {
        val hint2 = seek(r, key)
        hint2.copy(left = join(l, m, v, hint2.left))
  } } }
  def end(dict: D): H = FocusHint(dict, ord.max, mon.zero, empty())
  def isEnd(dict: D, hint: H): Boolean = hint == end(dict)
  def next(dict: D, hint: H): H = {
    val rightDict = after(hint)
    val nextHint = begin(rightDict)
    if(isEnd(rightDict, nextHint))
      end(dict)
    else {
      val (k, v) = current(nextHint)
      FocusHint(before(hint), k, v, after(nextHint))
  } }
  def deleteHint(dict: D, hint: H, k: K): D = append(hint.left, hint.right)
  def findHint(dict: D, hint: H, k: K): V = hint.value
}
\end{lstlisting}
\caption{The implementation of focus-based hinted dictionary.}
\label{fig:focusdict}
\end{figure}

\section{Concrete Implementations}
\label{sec:concrete}

In this section, provide three categories of concrete implementations for hinted dictionaries. 
We start by using sorted arrays as the underlying implementation in Section~\ref{sec:sortedarray}. 
Then, we show how unbalanced trees can be used for implementing hinted dictionaries in Section~\ref{sec:unbalancedtree}.
Finally, we use balanced trees as the representation for hinted dictionaries in Section~\ref{sec:balancedtree}.

\subsection{Sorted Array}
\label{sec:sortedarray}
Using sorted arrays is one \revision{of} the main techniques for a sequential implementation of ordered dictionaries. In the C++ world, the \code{flat_map} container provided by the Boost library~\cite{boostflatmap} uses sorted arrays for representing dictionaries.
This data structure is particularly useful for the workloads where the insertions are applied to the end of the ordered dictionary.

\begin{figure}[t!]
\begin{lstlisting}
import scala.collection.mutable.ArrayBuffer

case class ArrayView[T](buffer: ArrayBuffer[T], lower: Int, upper: Int)

trait ArrayDict[K, V] extends FocusHintedDict[K, V, ArrayView[(K, V)]] {
  type D = ArrayView[(K, V)]
  def empty(): D = ArrayView(ArrayBuffer(), 0, 0)
  def isEmpty(dict: D): Boolean = dict.upper == dict.lower
  def begin(dict: D): H = 
    if(isEmpty(dict)) end(dict)
    else {
      val (k, v) = dict.buffer(dict.lower)
      FocusHint(empty(), k, v, ArrayView(dict.buffer, dict.lower + 1, dict.upper))
    }
  def middle(dict: D): H = {
    if(isEmpty(dict)) end(dict)
    else {
      val mid = (dict.lower + dict.upper) / 2
      val (k, v) = dict.buffer(mid)
      val l = ArrayView(dict.buffer, dict.lower, mid)
      val r = ArrayView(dict.buffer, mid + 1, dict.upper)
      FocusHint(l, k, v, r)
} } }

class CopyingArrayDict[K, V](
  implicit override val ord: Orderable[K], val mon: Monoid[V]) extends ArrayDict[K, V]
  with InsertBasedDict[K, V, ArrayView[(K, V)], FocusHint[K, V, ArrayView[(K, V)]]] {
  def insertHint(dict: D, hint: H, key: K, value: V): D = {
    val array = dict.buffer.clone() // can be removed if dict is no longer used
    val idx = before(hint).upper
    val (prevKey, prevValue) = current(hint)
    val newUpper = 
      if(prevValue == mon.zero) {
        array.insert(idx, (key, value))
        dict.upper + 1
      } else {
        array(idx) = (key, mon.op(prevValue, value))
        dict.upper
      }
    ArrayView(array, dict.lower, newUpper)
} }
\end{lstlisting}
\caption{The implementation of hinted dictionaries using an underlying sorted array.}
\label{fig:sortedarray}
\end{figure}

Inspired by these C++ implementations, Figure~\ref{fig:sortedarray} provides the implementation of hinted dictionaries using sorted arrays.
In order to preserve an underlying array we use \code{ArrayBuffer}s, mutable containers similar to \code{std::vector}s of C++.
The \code{ArrayView} data type represents a subset of an \code{ArrayBuffer} bounded by the indices specified by \code{lower} and \code{upper}.

The implementations for \code{empty}, \code{isEmpty}, and \code{begin} are straightforward. 
For the \code{middle} method, we need to first retrieve the index of the middle element (\code{mid}). 
Then, we compute the preceding dictionary by using the same lower bound, but with the upper bound specified by \code{mid}.
Similarly, the succeeding dictionary uses the lower bound specified by \code{mid+1}, but with the same upper bound.
Finally, we return the focused hint object based on the key-value pair of the middle element and the preceding and succeeding dictionaries.

\revision{To create the left and right \code{ArrayView}s of the hint object, we can share the underlying buffer of the current \code{ArrayDict} without copying it. Since \code{ArrayView} is using \code{ArrayBuffer} as the underlying array and the start and end of the \code{ArrayView} are determined by \code{lower} and \code{upper}, changing these parameters can result in a new dictionary. This sharing opportunity frees the code from copying elements every time we are using \code{middle}. A similar opportunity is available for \code{before} and \code{after}.}

Because of using arrays as the underlying representation, it would be more efficient to follow an \code{InsertBasedDict} interface.
To implement the \code{insertHint} method, we need to check if the hint object points to an actual element. 
This is achieved by checking if the associated value is different than the identity element of the underlying monoid.
In the case of the existence of an element with the same key, the size of the underlying \code{ArrayBuffer} does not need to change; 
it is sufficient to update the value of the existing element by applying the binary operator of the monoid to the previous value and the new value to be inserted (\code{mon.op(prevValue, value)}).
If the hint object does not point to an actual element, we need to insert the given key-value pair in the specified position.
Finally, we adjust the upper bound and return the updated array.

Note that in the case that there is no more references to the input dictionary in the user program, one can perform in-place update and there would be no need to copy the original array. 
We leave the implementation of the in-place update version for the sake of brevity.

\begin{figure}[t!]
\begin{lstlisting}
sealed trait Tree[T]
case class Bin[T](l: Tree[T], v: T, r: Tree[T]) extends Tree[T]
case class Leaf[T]() extends Tree[T]

trait TreeDict[K, V, E] extends FocusHintedDict[K, V, Tree[E]] {
  type Entity = E
  type D = Tree[E]
  def key(e: Entity): K
  def value(e: Entity): V
  def empty(): D = Leaf()
  def isEmpty(dict: D): Boolean = dict == Leaf()
  def seekFirst(dict: D): (D, K, V) = {
    val FocusHint(l, k, v, r) = middle(dict)
    if(isEmpty(l)) (r, k, v)
    else {
      val (tp, kp, vp) = seekFirst(l)
      (join(tp, k, v, r), kp, vp)
  } }
  def begin(dict: D): H =
    if(isEmpty(dict)) end(dict)
    else {
      val (r, k, v) = seekFirst(dict)
      FocusHint(empty(), k, v, r)
    }
  def middle(dict: D): H = dict match {
      case Leaf() => end(dict)
      case Bin(l, e, r) => FocusHint(l, key(e), value(e), r)
    }   
}
\end{lstlisting}
\caption{The generalized implementation for tree-based representations of hinted dictionaries.}
\label{fig:treehinted}
\end{figure}

\subsection{Unbalanced trees}
\label{sec:unbalancedtree}
An alternative implementation for hinted dictionaries is based on a tree-based representation. 
We first give a generic implementation for tree-based hinted dictionaries that can be used for both unbalanced and balanced tree representations. 
Afterwards, we show a simple representation where no smart effort is invested for maintaining the tree in balance.

Figure~\ref{fig:treehinted} provides the generalized implementation for ordered dictionaries using trees.
The \code{Tree} data type is defined as \revision{an ADT (Algebraic Data Type)}, where \code{Leaf} corresponds to a leaf and \code{Bin} specifies an intermediate node.
As the tree nodes can maintain additional information (e.g., height for balanced trees),
the type member \code{Entity} is used for keeping the type of the information kept by each tree node.
The \code{key} and \code{value} methods are used to extract keys and values from the elements, respectively. 

As opposed to sorted arrays, the tree-based hinted dictionaries have a straightforward implementation for \code{middle}; for leaves the hint object for \code{end} is returned, whereas for intermediate nodes the focused-hint object with 1) the left sub-tree as the preceding dictionary, 2) the right sub-tree as the succeeding dictionary, and 3) the key/value of its element as the key-value pair is returned.
To implement \code{begin}, the helper method \code{seekFirst} is defined, which looks for the smallest key-value pair and returns them alongside the succeeding dictionary. 
These values are used to return the hint object with an empty preceding dictionary.

Figure~\ref{fig:unbalancedtree} shows the implementation for hinted dictionaries that use unbalanced trees.
Because of the tree-based representation, a natural interface for \code{UnbalancedDict} is the join-based hinted dictionary, although one could use the insert-based one with worse performance.
The tree nodes do not need to maintain any additional information. Thus, the entity type of the tree nodes is the key-value pair (\code{(K, V)}).
The implementation for the \code{join} is to simply create an intermediate node with the given key-value pair as the content, and first and section dictionaries as the left and right sub-trees.

\begin{figure}[t!]
\begin{lstlisting}
class UnbalancedDict[K, V](implicit override val ord: Orderable[K], val mon: Monoid[V]) 
  extends TreeDict[K, V, (K, V)] 
  with JoinBasedDict[K, V, Tree[(K, V)], FocusHint[K, V, Tree[(K, V)]]]
  with HintedDictFoldTree[K, V, Tree[(K, V)], FocusHint[K, V, Tree[(K, V)]]] {
  def key(e: Entity): K = e._1
  def value(e: Entity): V = e._2
  def join(left: D, key: K, value: V, right: D): D = Bin(left, (key, value), right)
}
\end{lstlisting}
\caption{The implementation of hinted dictionaries using unbalanced binary trees.}
\label{fig:unbalancedtree}
\end{figure}

\subsection{Balanced trees}
\label{sec:balancedtree}

Figure~\ref{fig:balancedtree} shows the generalized implementation for balanced-tree-based hinted dictionaries.
This interface subsumes AVL trees and WBB trees.
It is possible to implement Red-Black tree and Treaps as hinted dictionaries by appropriately overriding 
the \code{join} method, however, as it was shown that AVLs and WBBs have superior
performance in comparison with them~\cite{blelloch2016just}, we only present their implementation in this \paper. 

\begin{figure}[t!]
\begin{lstlisting}[backgroundcolor=\color{revisionbgcolor}]
trait BalancedDict[K, V, N] extends TreeDict[K, V, (K, V, N)] 
  with JoinBasedDict[K, V, Tree[(K, V, N)], FocusHint[K, V, Tree[(K, V, N)]]] {
  def key(e: Entity): K = e._1
  def value(e: Entity): V = e._2
  def info(e: Entity): N = e._3
  def zeroInfo: N
  def info(e: D): N = e match {
    case Leaf() => zeroInfo
    case Bin(_, entity, _) => info(entity)
  }
  def newInfo(left: N, right: N): N
  def rotateLeft(tree: D): D = tree match {
    case Bin(l, e1, Bin(l2, e2, r2)) =>
      bin(bin(l, key(e1), value(e1), l2), key(e2), value(e2), r2)
    case _ => throw new Exception("Not left rotatable")
  }
  def rotateRight(tree: D): D = // elided for brevity
  def isHeavier(left: D, right: D): Boolean
  def isVeryHeavier(left: D, right: D): Boolean
  def joinRight(left: D, key: K, value: V, right: D): D = 
    if(!isHeavier(left, right)) bin(left, key, value, right)
    else {
      val hint = middle(left)
      val (leftLeft, leftRight) = (before(hint), after(hint))
      val (k, v) = current(hint)
      val newRight = joinRight(leftRight, key, value, right)
      if(isHeavier(newRight, leftLeft)) {
        val newHint = middle(newRight)
        if(isVeryHeavier(before(newHint), after(newHint)))
          rotateLeft(bin(leftLeft, k, v, rotateRight(newRight)))
        else rotateLeft(bin(leftLeft, k, v, newRight))
      } else bin(leftLeft, k, v, newRight)
    }
  def joinLeft(left: D, key: K, value: V, right: D): D = // elided for brevity
  def bin(left: D, key: K, value: V, right: D): D = 
    Bin(left, (key, value, newInfo(info(left), info(right))), right)
  def join(left: D, key: K, value: V, right: D): D = 
    if(isHeavier(left, right)) joinRight(left, key, value, right)
    else if (isHeavier(right, left)) joinLeft(left, key, value, right)
    else bin(left, key, value, right)
}
\end{lstlisting}
\caption{The generalized interface for hinted dictionaries using balanced binary trees.}
\label{fig:balancedtree}
\end{figure}

The interface of \code{BalancedDict} accepts the extra parameter \code{N} for the extra information
kept by the tree nodes. For example, AVL trees store the height of the tree in each node and WBB trees store the size of the tree as the weight information.

In order to preserve correct bookkeeping information, the smart constructor \code{bin} is added.
This method invokes the abstract method \code{newInfo} in order to compute the updated information
based on the information of the sub-trees. This method needs to be overridden by concrete balanced
tree implementation choices.

The \code{join} method starts with checking if the tree is unbalanced towards either of its sub-trees.
If this is not the case, the smart construct \code{bin} is invoked to simply construct the new root node.
However, if either of the sides is heavier, an appropriate recursive method is invoked in order to
take care of possible rotations.

In Figure~\ref{fig:balancedtree} we show the implementation of \code{joinRight}, which is invoked when
the left sub-tree is heavier than the right sub-tree.
The implementation of \code{isHeavier} is again postponed to the concrete implementation of a balanced tree.
The rest of the implementation of \code{joinRight} mirrors a generalized version of what has been reported before
in~\cite{blelloch2016just}. 
An interesting case is when there needs to be a double rotation involved. This is \revision{checked by 
\code{isVeryHeavier},} which needs to be implemented by a concrete balanced tree implementation.

\begin{figure}[t!]
\begin{lstlisting}[backgroundcolor=\color{revisionbgcolor}]
class AVLDict[K, V](implicit override val ord: Orderable[K], val mon: Monoid[V]) 
  extends BalancedDict[K, V, Int] {
  type N = Int
  def zeroInfo: Int = 0
  def newInfo(left: N, right: N): N = math.max(left, right) + 1
  def isHeavier(left: D, right: D): Boolean = info(left) > info(right) + 1
  def isVeryHeavier(left: D, right: D): Boolean = info(left) > info(right)
}
class WBBDict[K, V](implicit override val ord: Orderable[K], val mon: Monoid[V]) 
  extends BalancedDict[K, V, Int] {
  type N = Int
  val ALPHA = 0.29
  val RATIO = ALPHA / (1 - ALPHA)
  val BETA = (1 - 2 * ALPHA) / (1 - ALPHA)
  def zeroInfo: Int = 1
  def newInfo(left: N, right: N): N = left + right - 1
  def isHeavier(left: D, right: D): Boolean = RATIO * info(left) > info(right)
  def isVeryHeavier(left: D, right: D): Boolean = 
    info(left) > BETA * newInfo(info(left), info(right))
}
\end{lstlisting}
\caption{The implementation for hinted dictionaries based on AVL and WBB trees.}
\label{fig:concretebalancedtree}
\end{figure}

Figure~\ref{fig:concretebalancedtree} shows the implementation of AVL and WBB trees using the generalized interface mentioned above. 
The AVL tree maintains the height of the tree as the extra information. 
The height of a new tree is computed by incrementing the maximum height of its sub-trees by one.
A sub-tree is heavier than another sub-tree when its height is more than an increment of the height
of the other sub-tree. 
And finally, a node needs double rotation only if the height of its left sub-tree is more than the height of 
its right sub-tree. 

The WBB tree considers the size of the tree (added by one) as the extra information, referred to as weight~\cite{adams1993functional}.
The updated weight is computed by adding the weight of sub-trees (decremented by one).
The \code{RATIO} and \code{BETA} parameters control whether rotation or double rotation need to be performed.
The values for these parameters specify the trade-off between the tree being in a perfect balance and
the number of re-balancing invocations. 
There were follow up research on fixing the balancing issues related to the parameter values originally suggested by Adams~\cite{hirai2011balancing,straka2011adams}. 
We use the parameters reported by~\cite{blelloch2016just}.

\begin{figure}[t!]
  \centering
  \includegraphics[width=\columnwidth]{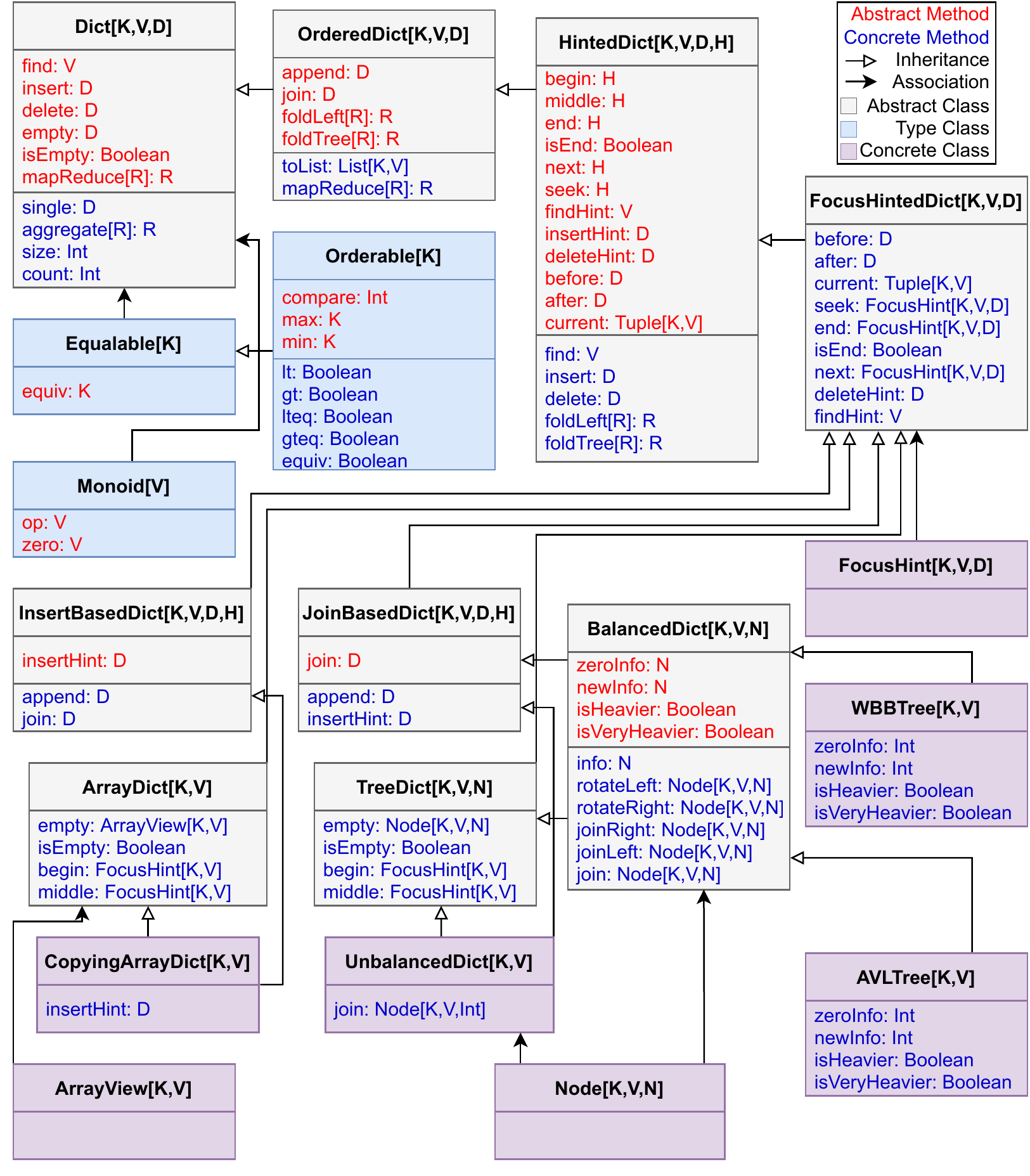}
  \caption{\revision{The bird's eye view of all the interfaces.}}
  \label{fig:putit}
\end{figure}

\revision{
\subsection{Putting It All Together}
\label{sec:putit}
Finally, we give an overall picture of hinted dictionaries by connecting all the pieces together. 
Figure~\ref{fig:putit} shows the different interfaces defined throughout this \paper.
To reduce the number of classes, we merged the definition of several interfaces with each other (e.g., \code{DictIterations} is merged with \code{Dict}). 
Crucially, there is no cake-pattern-based interface in Figure~\ref{fig:putit}, as this technique is not essential for implementing hinted dictionaries. 
}

\revision{

\section{Evaluation}
\label{sec:eval}
In this section, we evaluate the performance of hinted dictionaries. 
First, we present an efficient C++ prototype for them.
Then, we show the experimental results by comparing our C++ implementation with competitors for set-set and sparse vector operations. 

\subsection{Tuned Implementation in C++}
\label{sec:tuned}
The hinted dictionaries can have an efficient low-level implementation in C++. 
We provide the following hinted dictionary implementations: 1) \code{array_dict}, an array-based implementation that uses 
\code{std::vector<std::pair<K, V>>} as the underlying representation, and 2) \code{wbb_dict}, a tree-based implementation 
based on WBBs~\cite{wbtree,adams1993functional}.

We do not use the hierarchy presented in Figure~\ref{fig:putit} for performance reasons; we merge the definitions of all the parent interfaces of hinted dictionaries into \code{array_dict} and \code{wbb_dict}.
The implementation for \code{wbb_dict} stays very similar to the one presented in Section~\ref{sec:balancedtree}.
However, we have applied the following performance tuning tricks for \code{array_dict}, the interface of which is shown in Figure~\ref{fig:arrdictcpp}.
}

\begin{figure}[t!]
\begin{lstlisting}[language=C++,backgroundcolor=\color{revisionbgcolor}]
template<class K, class V, class Compare, class Mon>
class array_dict {
    private: 
        typedef std::vector<std::pair<K, V>> vector_t;
        typedef std::pair<int, bool> hint_t;
        int lower; int upper;
        vector_t buffer;
    public:
        void insert(K& key, V& value);
        hint_t insert_hint(hint_t& hint_obj, K& key, V& value);
        bool is_end(hint_t& hint_obj);
        template<SearchMethod search_method> 
        hint_t seek(K& key);
        void after(hint_t& hint_obj);
        template<class R, class Func> 
        void inplace_fold_left(R& z, Func op);
        /* Elided for brevity */
}
\end{lstlisting}
\caption{\revision{The interface of sorted-array-based dictionaries in C++.}}
\label{fig:arrdictcpp}
\end{figure}

\smartpara{Pointer-Based Hints} In Section~\ref{sec:focushint} we presented \code{FocusHint} objects that materialize the entire dictionaries before and after a hint object. 
However, this design results in unnecessarily copying of arrays (cf. Section~\ref{sec:sortedarray}). 
The \code{array_dict} implementation only maintains a pointer to the appropriate place by using pointer-based hint objects of type \code{std::pair<int, bool>}. 
If the hint object points to an actual element of the hinted dictionary, the first element of the pair specifies its index and the second element is set to \code{true}.
For keys that do not exist in the dictionary (i.e., the associate value is the zero element of the underlying monoid), 
the first element is the index of an element with the \textit{least upper bound} key and the second element is set to \code{false}.

\smartpara{Binary Search vs. Linear Search} In Section~\ref{sec:focushinteddict} we used binary search (by calling \code{middle}) in order to implement \code{seek}.
However, as we observe in the next section, in many cases it could be beneficial to use linear search. 
We provide a template parameter for the \code{seek} method in order to specify the underlying search method.

\smartpara{In-Place Updates} Finally, we use in-place updates to improve the performance in the following cases. 
First, the methods that return a subset of the hinted dictionary (\code{before} and \code{after}) can perform an in-place modification of the boundary of the dictionary (\code{lower} and \code{upper}). Second, the aggregation-based methods that produce dictionaries can use a single mutable dictionary and modify it in-place, instead of passing around new dictionaries (cf. \code{inplace_fold_left} in Figure~\ref{fig:arrdictcpp}).

\smartpara{Constant-time Size} The \code{array_dict} implementation can compute the size of the dictionary by evaluating \code{upper - lower}. 
Similarly, \code{wbb_dict} can compute the size using the meta data (\code{info}).
However, both dictionaries still require iterations for \code{count} (i.e., the number of elements with a non-zero value).
Thanks to the fast size computation, we can make sure that we always iterate over smaller dictionaries for all binary operations over dictionaries (e.g., set-set operations).

\subsection{Experimental Results}
\label{sec:exp_res}
\smartpara{Experimental Setup}
We run our experiments on a machine running Ubuntu 20.04.3 equipped with
an Intel Core i5 CPU running at 1.6GHz, 16GB of DDR4 RAM. We use G++ 9.4.0 for compiling the generated C++ code using the O3 flag. We also use SciPy 1.8.0 (on Python 3.8.10) as the competitor.

\smartpara{Workloads}
We consider the following set-set and sparse vector operations: 1) set-set union, 2) set-set intersection, 3) sparse vector addition, 4) sparse vector element-wise multiplication, and 5) sparse vector inner product.
For all the experiments, we generate randomly synthetic data by varying the size of sets and the density of vectors. We run all the experiments for ten times and measure their average run time.

\smartpara{Competitors}
We consider the following alternatives of our C++ prototype and other frameworks:

\begin{itemize}
\item \textbf{array\_dict (Linear):} array-based dictionary with linear-search-based seek.
\item \textbf{array\_dict (Binary):} array-based dictionary with binary-search-based seek.
\item \textbf{wbb\_dict:} tree-based dictionary that uses a WBB-based representation.
\item \textbf{Baseline C++:} a baseline implementation using the operations provided by \code{std::set} (for the set experiments) or \code{std::map} (for the sparse vector experiments).
\item \textbf{std::set\_intersect, std::set\_union:} set-set operations provided by the standard library of C++. As input arguments we use \code{std::set} collections.
\item \textbf{SciPy:} sparse linear algebra operators provided by the SciPy library. A sparse vector is represented as a row matrix with a CSR (Compressed Sparse Row) or a column matrix with a CSC (Compressed Sparse Column) format.

\end{itemize}

\begin{figure*}[t]
\includegraphics[width=0.49\textwidth]{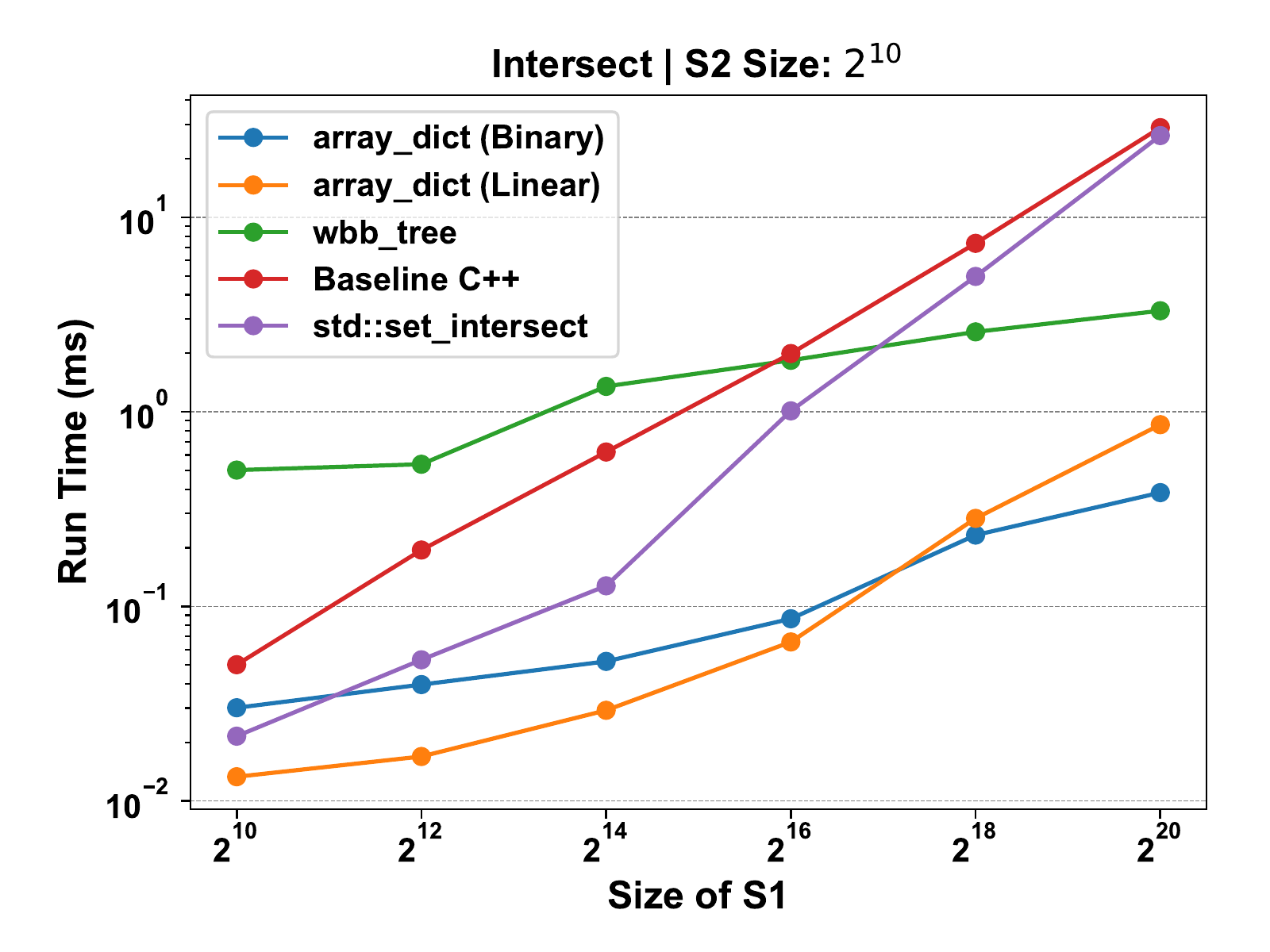}~%
\includegraphics[width=0.49\textwidth]{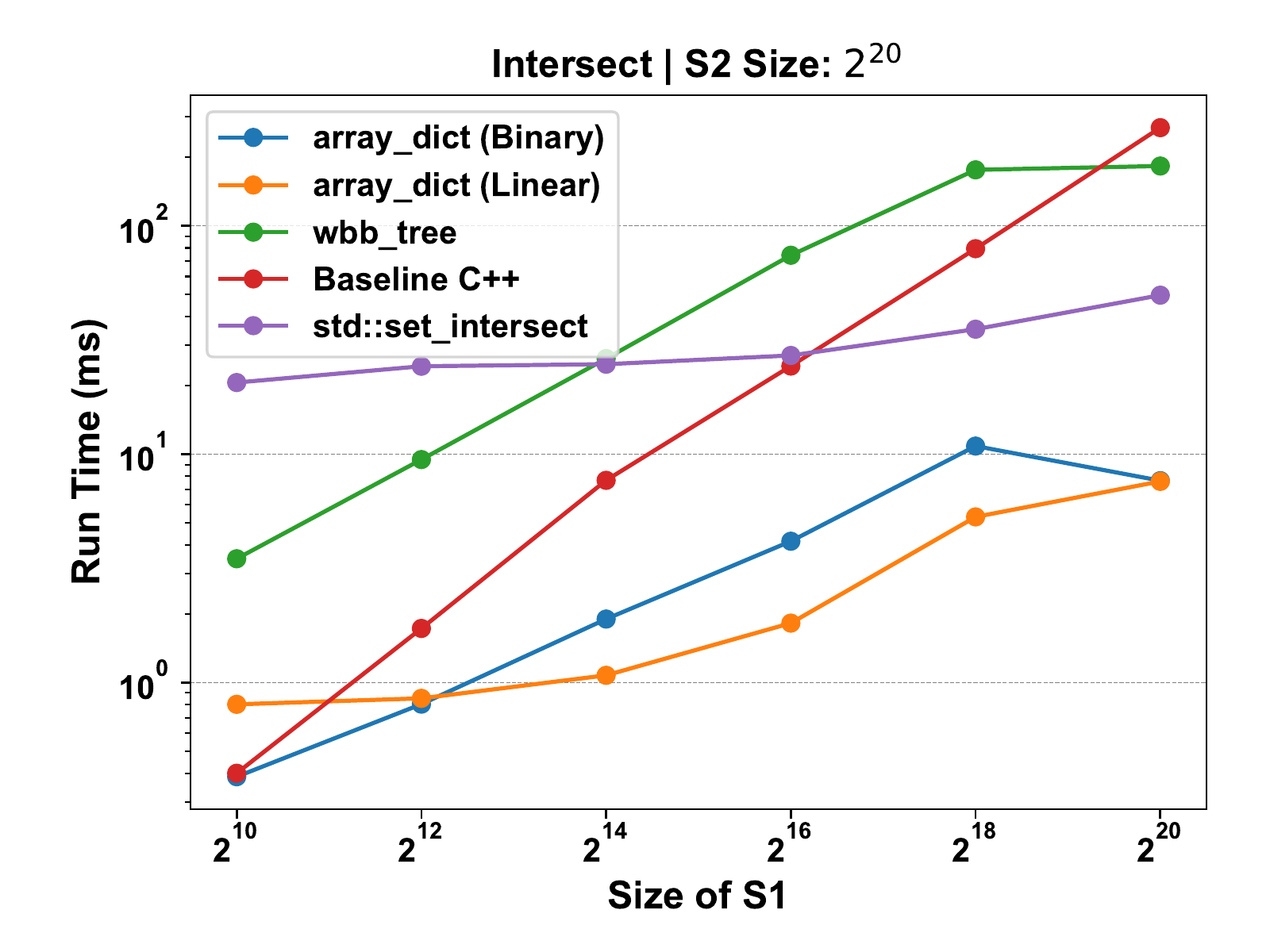}

\includegraphics[width=0.49\textwidth]{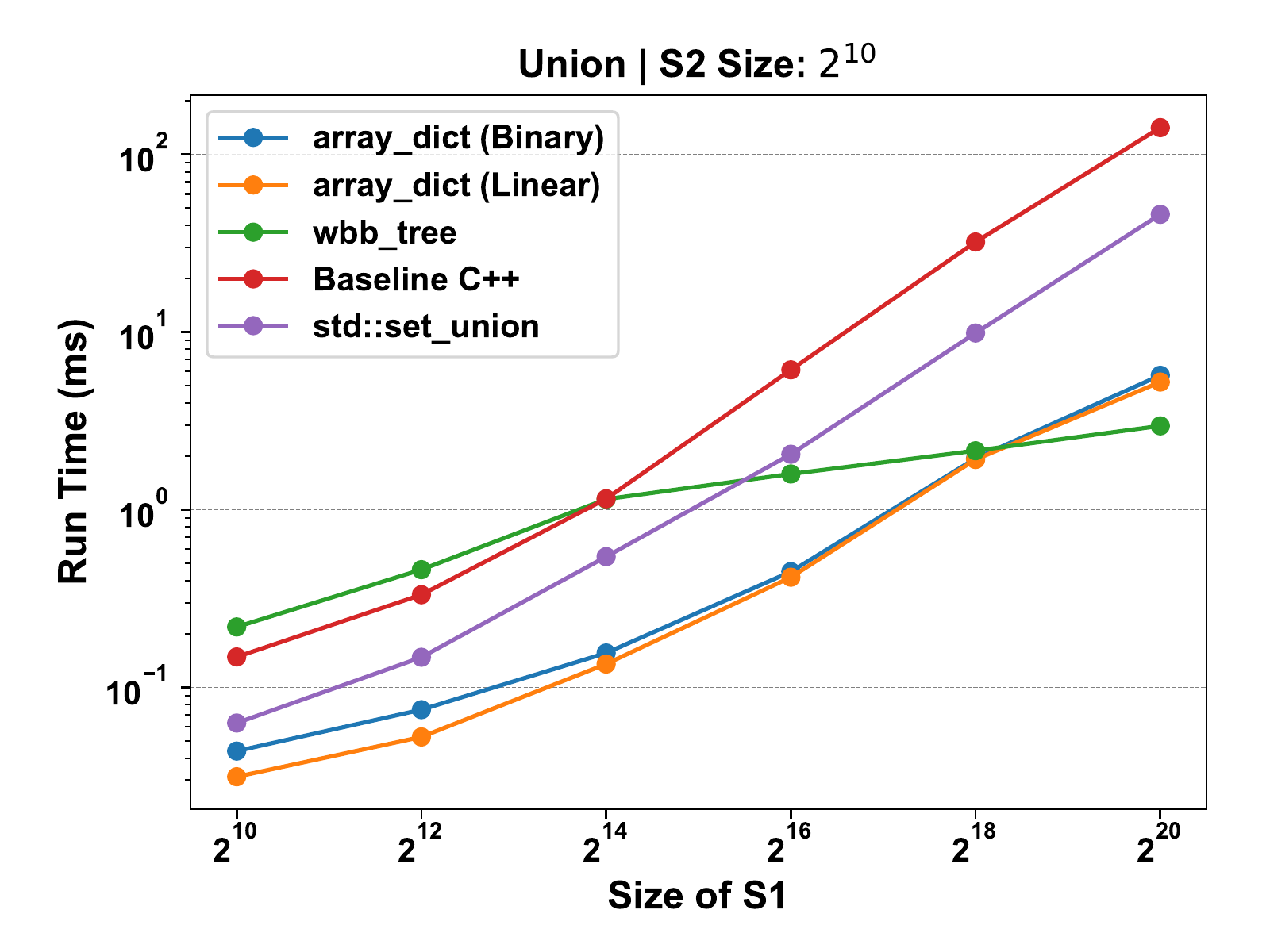}~%
\includegraphics[width=0.49\textwidth]{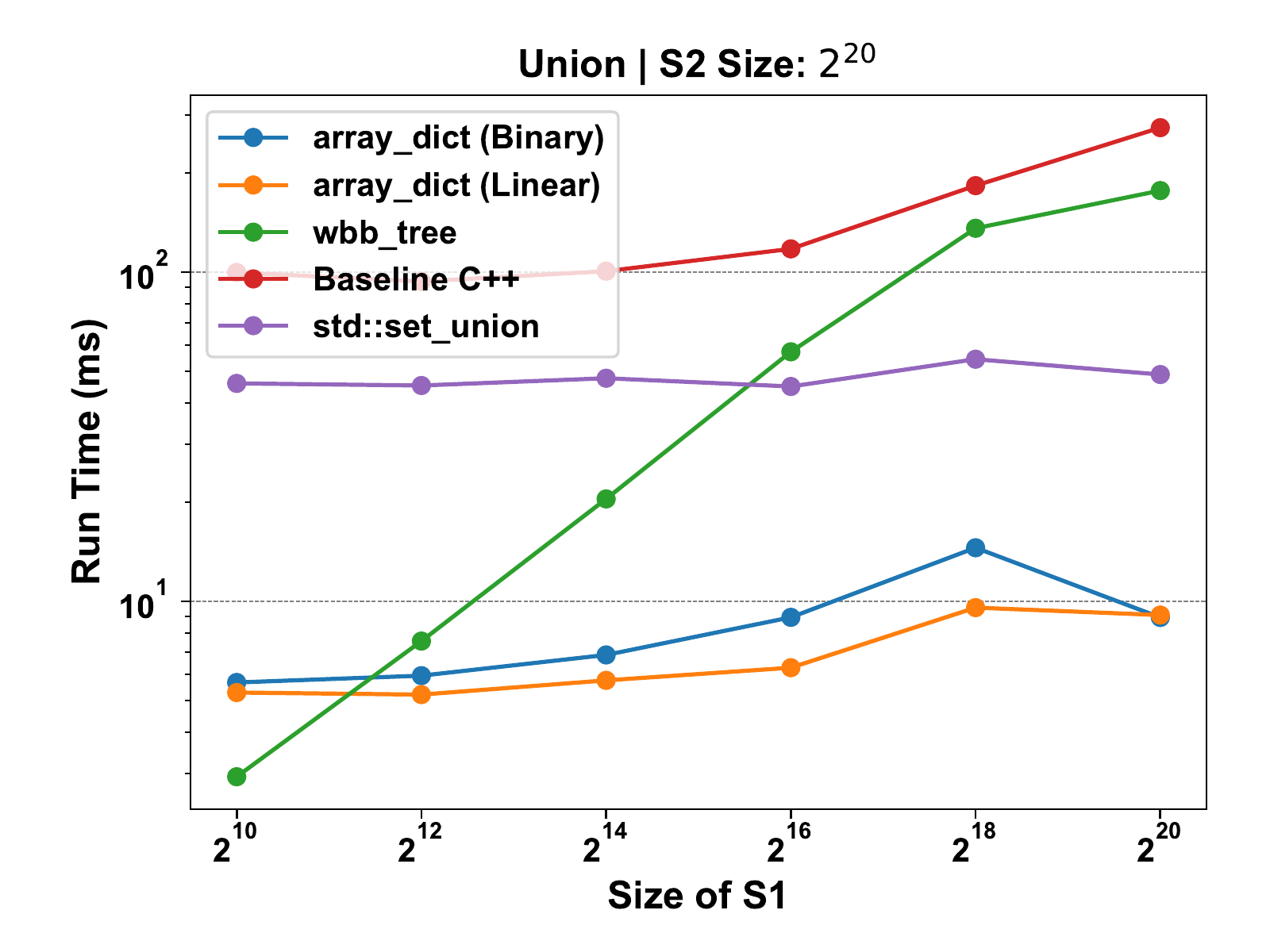}
\vspace{-0.4cm}
\caption{\revision{Experimental results for set-set operations.}}
\label{fig:exp:la}
\vspace{-0.5cm}
\end{figure*}

\smartpara{Set-Set Experiments}
Figure~\ref{fig:exp:la} shows the results for the union and intersection operations over sets. We make the following observations.
First, in most cases, we observe a superior performance for the \code{array_dict} implementations.
In most cases, the linear-search-based approach has better performance.
However, as the difference between the size of sets widens, the binary-search starts showing better performance.
This is because of the additional run-time improvements caused by skipping irrelevant elements.

Furthermore, we observe that the \code{wbb_dict} implementation does not show superior performance in most cases. 
As shown before~\cite{blelloch2016just}, one of the advantages of such join-based implementations is their amenability to parallelism that we leave for the future. 

Finally, we observe an asymptotically improved performance over the baseline C++ implementation. This is thanks to turning $O(\log n)$ lookup and insertion operations into amortized constant-time ones. 
The implementations provided by the standard library of C++ suffer from a similar issue. 
In addition, due to a concurrent linear iteration over both sets, they show worse performance for the sets with a large difference in their sizes.

\begin{figure*}[t]
\includegraphics[width=0.49\textwidth]{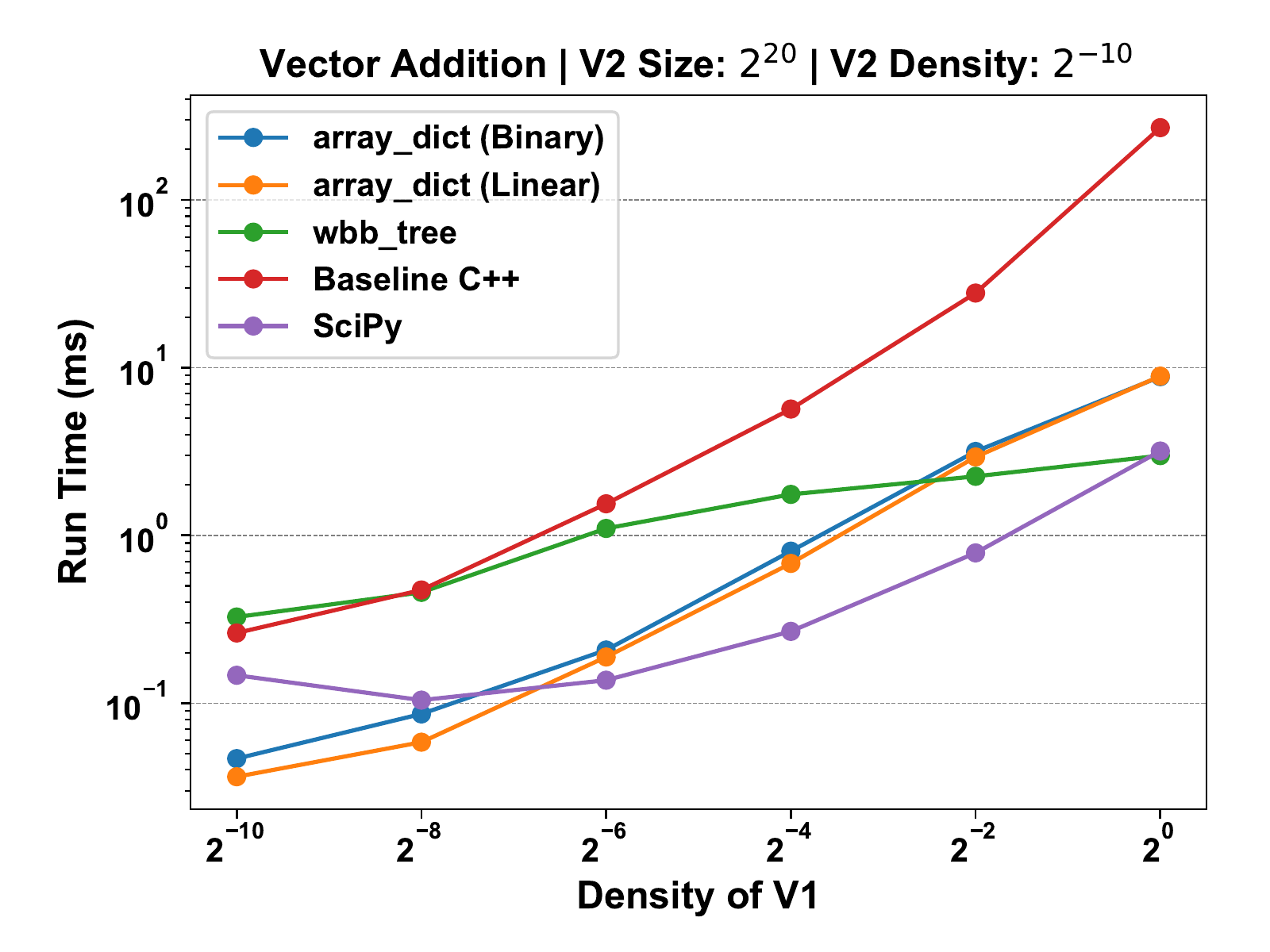}~%
\includegraphics[width=0.49\textwidth]{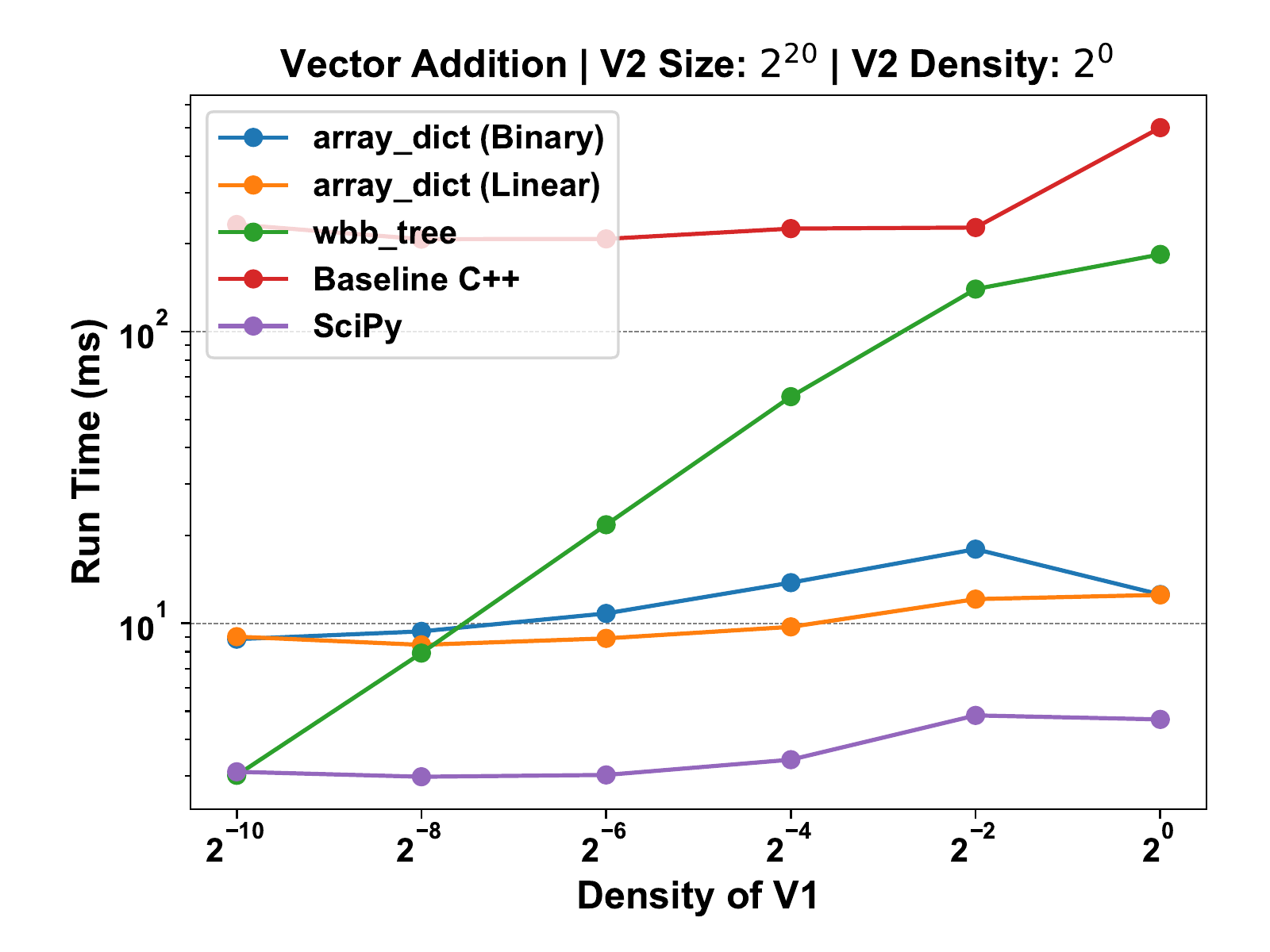}

\includegraphics[width=0.49\textwidth]{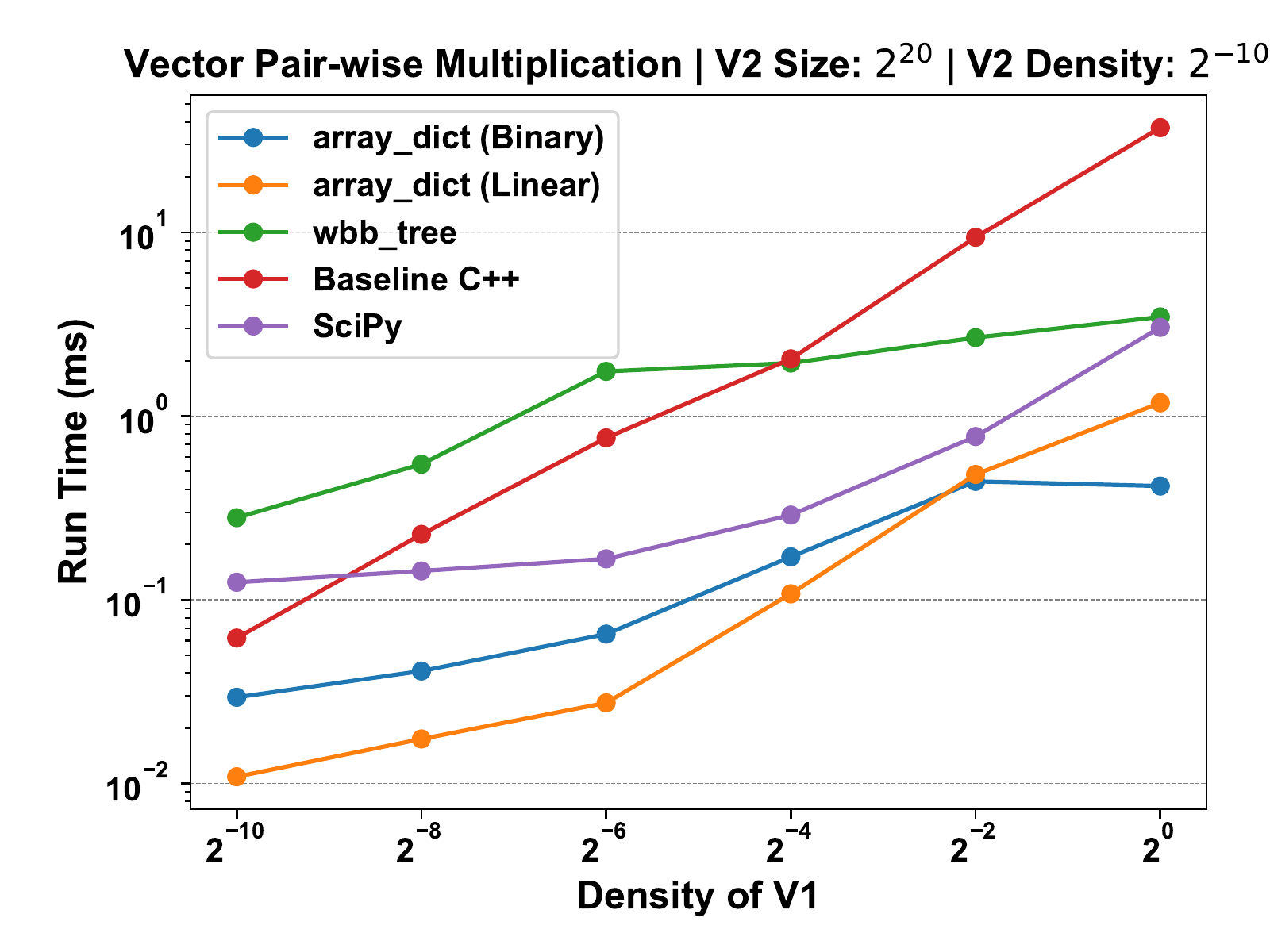}~%
\includegraphics[width=0.49\textwidth]{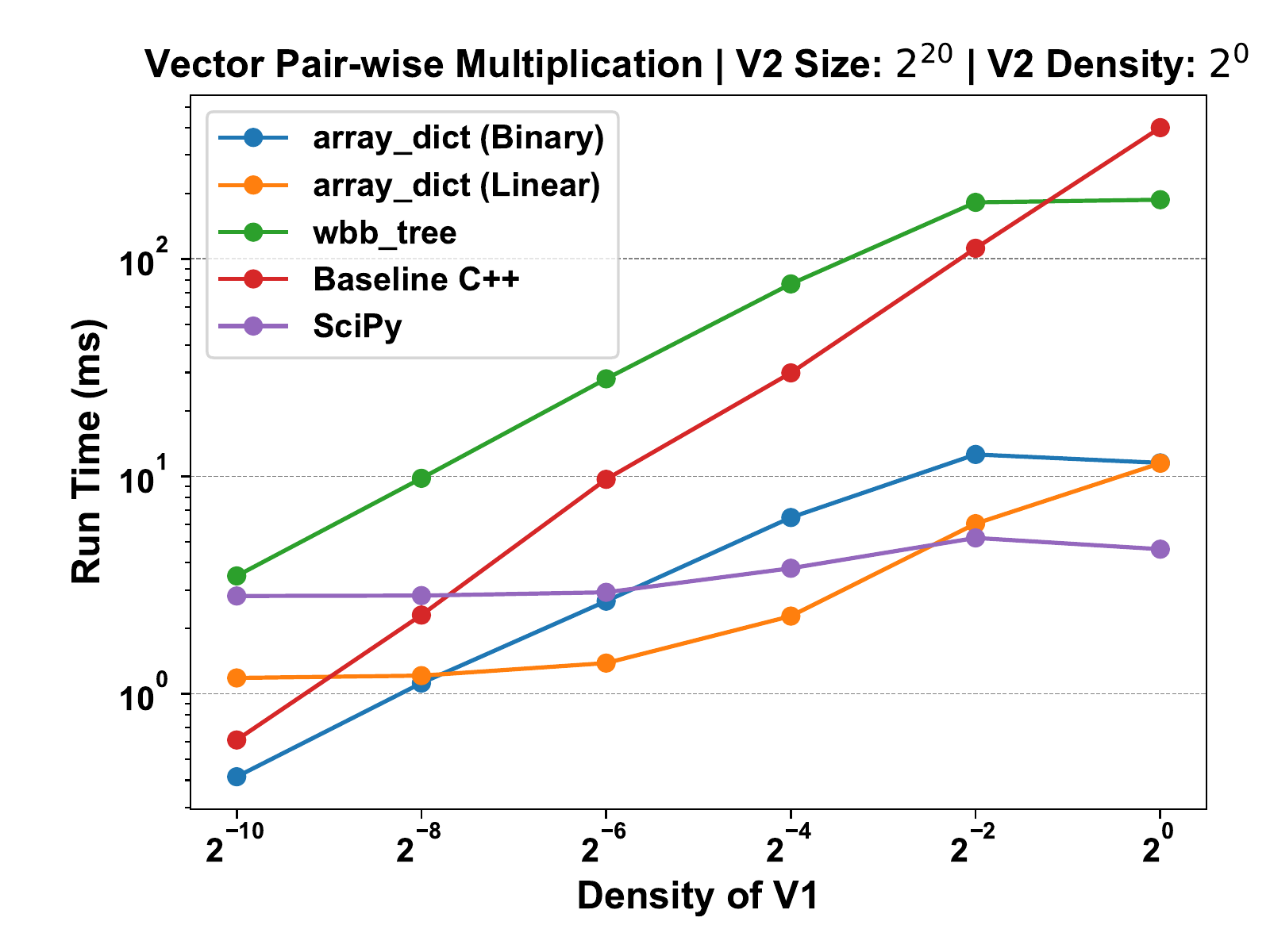}

\includegraphics[width=0.49\textwidth]{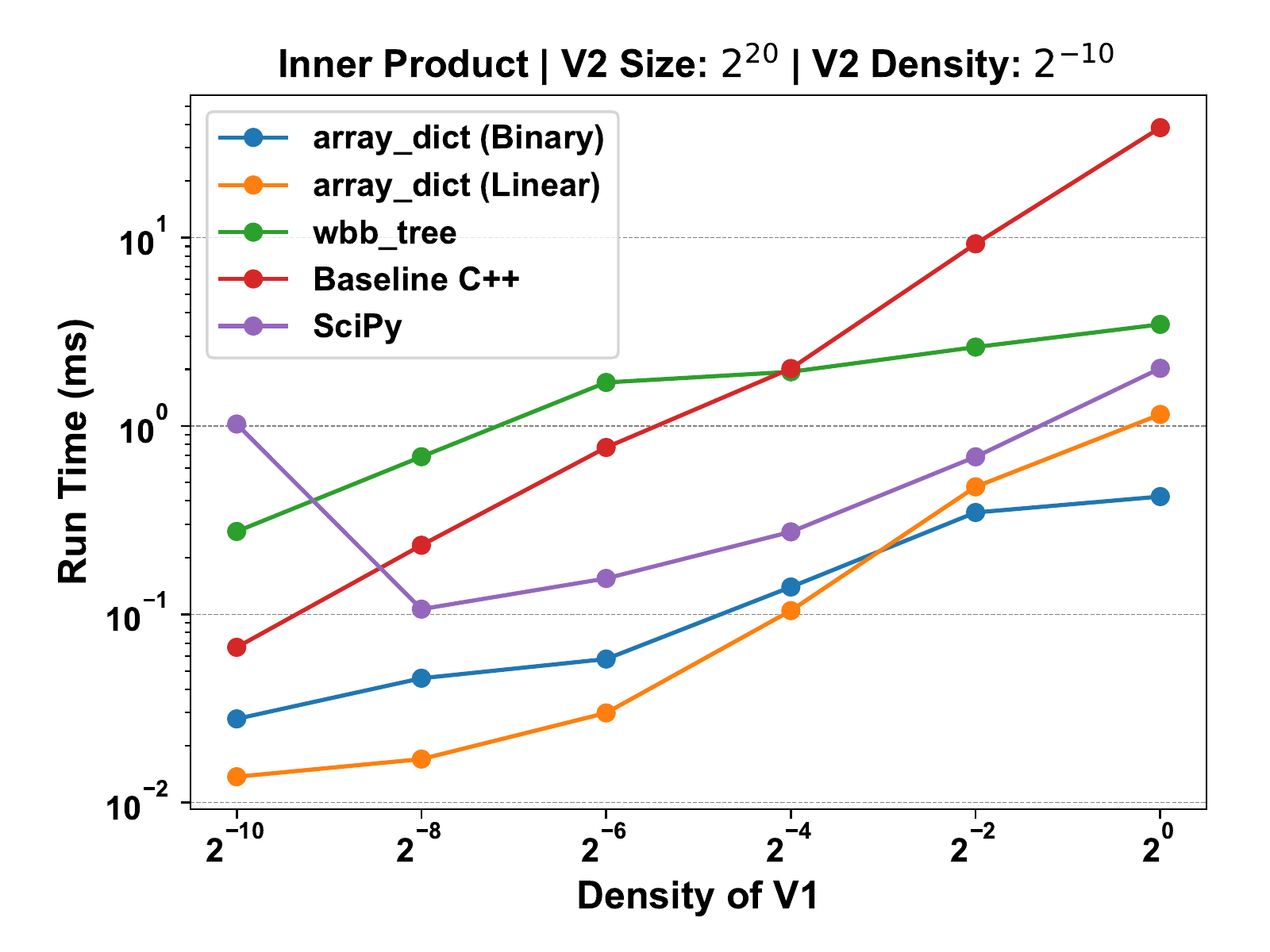}~%
\includegraphics[width=0.49\textwidth]{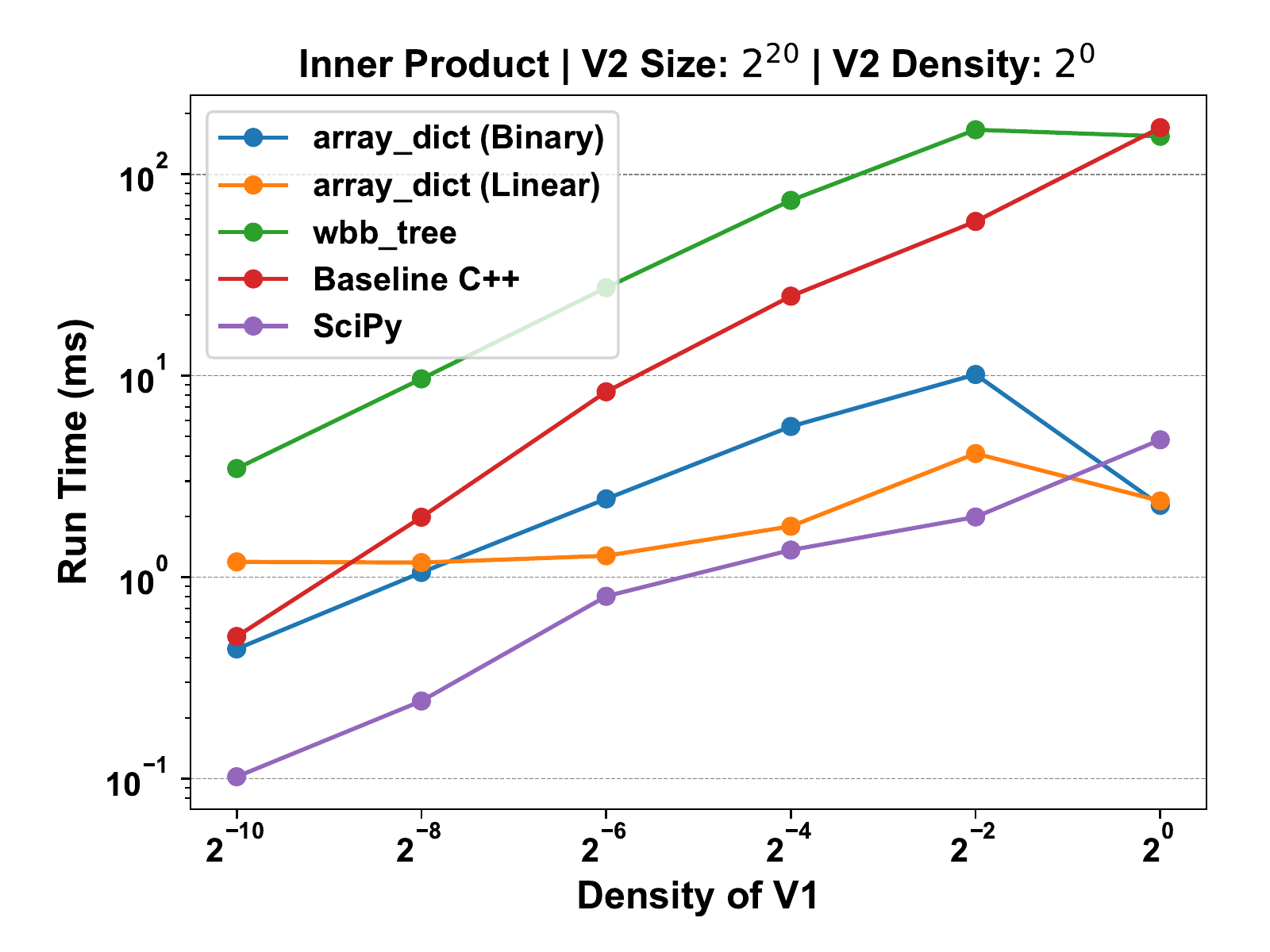}
\vspace{-0.4cm}
\caption{\revision{Experimental results for sparse vector operations.}}
\label{fig:exp:la}
\vspace{-0.5cm}
\end{figure*}

\smartpara{Sparse Vector Experiments} The implementations for sparse vector addition and element-wise multiplication are identical to the ones for set union and intersection, respectively.
The sparse vector addition uses real number addition instead of boolean disjunction, and the element-wise multiplication uses real number multiplication instead of boolean conjunction.
The results for vector inner product are also very similar to the ones for the element-wise multiplication (cf. Figure~\ref{fig:exp:la}).

The SciPy framework uses a CSR format for representing all vectors, except for the second operand of the vector inner product. 
This is because the second vector needs to be transposed which makes the CSC format a better representation. Overall, this framework shows superior performance for vectors with a large density. 
This can be related to their better storage layout (struct of array instead of array of struct) that leads to improved cache locality.
The \code{array_dict} variants show better performance for vectors with a higher degree of sparsity.

\section{\revision{Conclusion and Outlook}}
\label{sec:concl}

In this \paper, we introduced hinted dictionaries, a unified technique for implementing ordered dictionaries.
We have shown how hinted dictionaries unify the existing techniques from both imperative and functional languages.
These dictionaries can be used as the collection type for data-intensive workloads.
It would be interesting to see the usage of such data structures for real-world use-cases such as query processing (relying on relations in the form of sets and bags) as well as sparse linear algebra (relying on sparse vectors and matrices).

\revision{
The performance improvement offered by hinted dictionaries does not come for free.
The programmers must be careful on how to use hinted dictionaries.
As presented in Section~\ref{sec:hinteddict},
certain preconditions need to be preserved for hint objects.
Violating these preconditions by the programmers destroys the invariants of hinted dictionaries leading to runtime errors or, even worse, undefined behaviour, 
which can hinder the productivity of programmers. 
One interesting future direction is to statically detect the violation of the hinted dictionaries' invariants.

Furthermore, we envision the following future directions for hinted dictionaries.
First, we plan to consider real-world applications that require batch processing of sets and
maps, including relational query engines and sparse tensor processing frameworks.
Furthermore, it would be interesting to use code generation and multi-stage programming techniques to generate low-level code. 
This way, one can automatically improve the performance by removing allocation of unnecessary intermediate objects (e.g., \code{FocusHint} objects) or to use in-place updates (cf. Section~\ref{sec:tuned}) from a purely functional implementation.
Finally, for applications such as query processing the trade-offs between hashing and sorting have been debated for a long time. We believe hinted dictionaries provide a nice abstraction layer for DSLs based on dictionaries (e.g., the physical query plan of query engines) to automatically tune the choice of the underlying dictionary implementation.
}

\section*{Acknowledgements}
The authors would like to thank Huawei for their support of the distributed data management and
processing laboratory at the University of Edinburgh.

\bibliography{refs} 

\end{document}